\documentclass[12pt]{article}
\usepackage{amssymb}
\usepackage{amsmath}
\usepackage{amsthm}
\usepackage{color}
\usepackage{bm}
\usepackage{natbib}
\usepackage{xcolor}
\usepackage{graphicx}
\usepackage{booktabs}
\usepackage{multirow}
\usepackage{soul}
\usepackage{float}
\restylefloat{table}
\usepackage{url}
\usepackage{pdfpages}

\usepackage{setspace}
\def\bbeta{{ \bm{\beta} }}
\def\vX{{ \bm{X} }}

\def\Tstardbeta{{T^{\ast}\big(d_{\bbeta}\big)}}
\def\Tstardbetai{{T^{\ast}_{i}\{d_{\bbeta}\big(\vX_{i}\big)\}}}

\def\Tstardx{{T^{\ast}\big(d(\vX)\big)}}
\def\Tstard{{T^{\ast}\big(d\big)}}

\def\TstarOne{{T^{\ast} \big(1\big)}}
\def\TstarZero{{T^{\ast}\big(0\big)}}
\def\TstarOnei{{T^{\ast}_{i} \big(1\big)}}
\def\TstarZeroi{{T^{\ast}_{i}\big(0\big)}}

\def\dbetax{{d_{\bbeta}\big( \vX \big)}}
\def\dbetaxi{{d_{\bbeta}\big( \vX_{i} \big)}}

\def\mathbbBo{{\mathbb{B}^o}}

\def\Qtaub{{Q_{\tau}\big\{T^{\ast}\big(d_{\bbeta}\big)\big\}}}

\def\Qtaubetax{{Q_{\tau}\big\{T^{\ast}\big(d_{\bbeta}(\vX)\big)\big\}}}
\def\QHattaubetax{{\widehat{Q}_{\tau}\big\{T^{\ast}\big(d_{\bbeta}(\vX)\big)\big\}}}

\def\Rbetai{{R_{i}\big(\bbeta\big)}}

\def\Gtrue{{G_{C}}}

\def\Ghat{{\widehat{G}_{C}}}

\newcommand{\TDOneNo}{{\dot{T}\left(d_1, \emptyset\right)}}
\newcommand{\TDOneDtwo}{{\dot{T}\left(d_1, d_2\right)}}

\newtheorem{theorem}{Theorem}
\newtheorem{lemma}{Lemma}

\def\Xbeta0X{{\big( -\tilde{\vX}^T \widetilde{\bbeta}_{0},\tilde{\vX}^{{T}} \big)^{{T}}}}

\usepackage{xcite}
\usepackage{xr}
\makeatletter
\newcommand*{\addFileDependency}[1]{
	\typeout{(#1)}
	\@addtofilelist{#1} 
	\IfFileExists{#1}{}{\typeout{No file #1.}} 
}
\makeatother

\begin{document}

\title{\large
\bf Transformation-Invariant Learning of Optimal Individualized Decision Rules with Time-to-Event Outcomes}
\author{Yu Zhou, Lan Wang, Rui Song and Tuoyi Zhao}
\date{}

\maketitle

\begin{singlespace}
\begin{footnotetext}[1]
{  Yu Zhou is a machine learning engineer at Roku. Email: izhou@roku.com. 
	Lan Wang is Professor, Department of Management Science, University of
	Miami. Email: lanwang@mbs.miami.edu.
	 Rui Song is Professor, Department of Statistics, North Carolina State University.
	 Email: rsong@ncsu.edu. Tuoyi Zhao is a Ph.D. student, Department of Management Science, 
	 University of Miami.  
	Wang, Zhou and Zhao's research was supported by NSF FRGMS-1952373. Song's research was supported by NSF DMS-2113637.
	We thank Co-Editor Professor McKeague, the AE and the anonymous referees for their helpful comments.
	}
\end{footnotetext}
\end{singlespace}

\begin{singlespace}
\begin{abstract}
	In many important applications of precision medicine, the outcome of interest is
	time to an event (e.g., death, relapse of disease)
	and the primary goal is
	to identify the optimal individualized decision rule (IDR) to prolong survival time.
    Existing work in this area have been mostly focused on estimating the optimal IDR to maximize the 
    restricted mean survival time in the population.  
	We propose a new robust framework for estimating an optimal static or dynamic 
   IDR with time-to-event outcomes based on an easy-to-interpret quantile criterion. 
   The new method does not need to specify an outcome regression model
   and is robust for heavy-tailed distribution.
   The estimation problem corresponds to a nonregular M-estimation problem with both finite and infinite-dimensional
   nuisance parameters. 
Employing advanced empirical process techniques, we establish the statistical theory of the estimated
parameter indexing the optimal IDR.
Furthermore, we prove a novel result that the proposed approach can consistently estimate the 
optimal value function under mild conditions even 
when the optimal IDR is non-unique, which happens in the 
challenging setting of exceptional laws. We also propose a smoothed resampling procedure
for inference.
The proposed methods are implemented in the R-package \texttt{QTOCen}. 
We demonstrate the performance of the proposed new methods via extensive Monte Carlo studies and 
a real data application.
\end{abstract}
\end{singlespace}


\newpage

\section{Introduction}
The problem of estimating the optimal individualized decision rule (IDR) has recently received substantial attention in precision medicine and other domains. 
A treatment can be a drug, a therapy, or any
other actionable choice (or a sequence of such choices) such as a policy
or program. 
The goal of optimal IDR estimation is to determine a decision rule that assigns a subject to one of the treatment options based on individual information available each decision point such that some functional 
of the potential outcome distribution is optimized.

For completely observed data, 
several successful approaches exist for estimating the optimal IDR,
including Q-learning \citep{watkins1992, Murphy05, chak2010, Song2015}, 
A-learning \citep{robins2000, Murphy2003, murphy2005experimental, moodie2010}, 
model-free or policy search methods
\citep{Robins2008, Orellan10, zhang2012robust, zhao2012estimating, zhao2015},
the interpretation-enhanced tree or list-based methods
\citep{laber2015tree, cui2017tree, zhu2017greedy, Zhang2017Lists} among others.
See also the books of \citet{CM2013} and \citet{KM2016} for a general introduction and additional references. 

The focus of this paper is on estimating the optimal static (one-stage) or dynamic (multi-stage) IDR 
for time-to-event or survival data, where the outcome is possibly
censored. Several new challenges arise when analyzing such data compared with the complete data case.
Censoring occurs when an individual drops out from the study or the study ends before 
the subject experiences the event of interest. 
The distribution of survival time (e.g., time to death, onset of disease) is often highly skewed. 
The situation gets even more complicated for estimating the optimal dynamic IDR, where
the data are collected longitudinally.
Consider the setting where the treatment decisions for a patient are made at $k$ pre-specified
decision points. Each decision is allowed to depend on the patient's characteristics
(e.g., gender, age) and treatment history (e.g., disease progression status and
how the individual responds to previous treatments) up to that decision point.
The patient may be censored at any stage of treatment. 
Direct application of existing complete-data
techniques could result in severe bias, as demonstrated in the Monte
Carlo studies in Section 5.

Several authors have recently investigated estimating the optimal
IDR with survival data, see 
\cite{goldberg2012q}, \cite{xu2016bayesian}, \cite{jiang2017estimation}, \cite{jiang2017doubly},  
\cite{bai2017optimal}, \cite{hager2018optimal}, \cite{xu2016bayesian},  
\cite{diaz2018targeted}, \cite{simoneau2019}, among others.
For time-to-event outcomes, new criterion is needed to 
evaluate the effectiveness of an IDR. 
Of crucial importance is that such a criterion can be reliably estimated under censoring. 
The existing work have been mostly focused on maximizing the 
{\it restricted mean survival time} in the population. 

We consider time-to-event outcomes and propose a new robust framework for estimating the optimal 
IDR using an alternative criterion based on the marginal quantile of the potential outcome
distribution.  
The new optimality criterion 
is easy to interpret.
Median survival time has already been popularly used 
clinically to evaluate the success of cancer treatment. 
It can be reliably 
estimated even under relatively heavy censoring (e.g., the censoring rate is more 
than 50\% in the real data example of this paper).
The resulted optimal IDR is invariant under monotone-transformation of the outcome. 
We develop robust estimation methods for both static and dynamic 
optimal IDRs. The robust approach 
circumvents the difficulty of specifying a reliable outcome regression model, 
especially for the dynamic setting which demands a sequence of generative regression models.

We consider estimating the optimal IDR in a class of candidate decision rules indexed by a finite-dimensional parameter.
The estimation problem corresponds to a challenging non-regular M-estimation
problem with both finite and infinite-dimensional nuisance parameters, due to the 
unknown censoring distribution. 
For the optimal static IDR, we rigorously establish the cube-root convergence rate for 
the estimated parameter by employing modern empirical processes
techniques. We prove that its asymptotic 
distribution corresponds to the maximizer of a centered Gaussian process
with a parabolic drift. It is worth emphasizing that the
nonstandard asymptotics is due to the intrinsic nature of the decision problem, 
which relates to a
sharp edge effect in the decision function. Due to the nature of nonstandard asymptotics, 
the problem is substantially harder than regular M-estimation problem with 
infinite-dimensional nuisance parameters. 

Moreover, we establish a useful novel result that shows the optimal value
can be consistently estimated under weak conditions
for the challenging setting of exceptional laws.
Under exceptional laws, there exists a subgroup of patients for whom the
treatment is neither beneficial nor harmful. In this case,
the optimal IDR is non-unique, see for example, the discussions in 
\cite{robins2014discussion} and \cite{Luedtke2016}.

Theoretically, our work complements existing results and significantly enhances the
knowledge about optimal IDR estimation with survival data.
The existing results have been mostly focused on prediction error bounds
and have not studied the properties of the estimated parameter indexing the
optimal IDR. Furthermore, to the best of our knowledge, existing work on optimal IDR estimation with survival outcomes
assume non-exceptional laws and thus avoid the problem of optimal
value estimation under exceptional laws.

The rest of the paper is organized as follows. Section 2 introduces the new framework and the robust method for estimating an optimal static IDR with survival data, with theoretical properties developed in Section 3. Section 4 presents the estimation method and the theory for the dynamic IDR setting. In Section 5, we report results from extensive Monte Carlo studies. In Section 6, we illustrate the application on the analysis of a breast cancer data set. The proposed methods are implemented in the R-package \texttt{QTOCen}. 
The regularity conditions are given in the Appendix. The technical derivations and additional numerical results are given in the online supplement.

\section{Robust estimation for static optimal IDR }
\subsection{Preliminaries}
We first consider the single-stage setting. 
Let $ A \in \{0, 1\}$ denote the binary treatment,
$ \vX \in \mathbb{R}^p $ denote the vector of covariates with support $ \mathcal{X}$,
and $ T\in \mathbb{R}^{+} $ denote the time to the event of interest (or a transformation thereof).
We often refer to $T$ as {\it survival time}.
Without loss of generality, we assume that a larger value of $T$ indicates better treatment effect.
The outcome $T$ may not be observed due to censoring. 
Let $C \in \mathbb{R}^{+}$ denote the censoring variable and 
$\Delta = I (T \leq C )$. If the observation is censored (i.e., $\Delta=0$), then we only observe 
$C$. Let $Y=\min\{T, C\} $ be the observed outcome.
The observed data consist of $\big\{\vX_i,A_i,Y_i,\Delta_i \big\}$, $i=1, \ldots, n$,
which are independent copies of $ \big(\vX,A,Y,\Delta \big) $.

To assess the treatment effect, we adopt the potential outcome framework \citep{neyman1923applications, rubin1978bayesian} in causal inference.
The $i$-th subject has two potential outcomes: $ \TstarZeroi$ and $\TstarOnei $,
where $T_i^*(0)$ is the survival time had the subject received treatment 0 and 
$T_i^*(1)$ is defined similarly, $i=1,...,n$. In practice, a subject receives one and only one of the two possible treatments.
Under the stable unit treatment value assumption (\citet{Rubin86}), 
we have
$
T_i=A_i\TstarOnei+\big(1-A_i\big) \TstarZeroi.
$
That is, the survival time of the $i$th subject is the potential survival time 
corresponding to the treatment that subject actually received.
Furthermore, we assume the no unmeasured confounders assumption \citep{RR83} is satisfied, i.e., 
$\{T_i^*(0), T_i^*(1)\}$ are independent of $A_i$ conditional on $\vX_i$. This is a common
assumption in causal inference and is automatically satisfied for a randomized trial.
Mathematically,
an IDR $ d(\vX) $ is a mapping from the space of covariates $ \mathcal{X} $ to the space of candidate treatments 
$\{0,1\}$. 
The potential outcome associated with $ d(\vX) $
is denoted by $\Tstardx$. We have
$
\Tstardx =\TstarOne d\big( \vX\big) +
\TstarZero \big[1-d\big( \vX\big)\big]. 
$


\subsection{New optimality criterion for evaluating IDR with time-to-event outcome} \label{subsection: new_criterion}

Different from the completely observed data case, the mean of the outcome is usually 
difficult to estimate accurately at the presence of censoring. 
The most popular criterion in the existing literature
for time-to-event data is the restricted mean survival time $\mbox{E}\{ \min \big(\Tstardx, L\big)\}$, where $L$ is a user-supplied cut-off time. 
In this paper, we consider an alternative criterion for comparing IDRs
for time-to-event outcomes based on the marginal quantile treatment effect.
The new criterion enjoys three appealing properties: easy to interpret, robust to long-tailed survival distribution, and invariant to the monotone transformation of survival time. The marginal $\tau$-th quantile ($0<\tau<1$) of 
the potential outcome $\Tstardx$ is defined as
\begin{equation*}
Q_{\tau}\big\{\Tstardx\big\} = \inf \Big\{ t\in\mathbb{R}:  P \big\lbrace \Tstardx \leq t \big\rbrace \geq \tau\Big\},
\end{equation*}  
where $P$ denotes the marginal distribution of $\Tstardx$. In IDR estimation problems,
$Q_{\tau}\big\{\Tstardx\big\}$ is referred to as the {\it value function}. 
We sometimes use the short-hand notation $ Q_{\tau}\{T^{\ast}(d)\}$. 
Given a class $ \mathcal{D}$ of candidate IDRs, the optimal IDR is defined as 
\begin{equation} \label{eq:general dopt}
d_{opt}(\Tstardx)=\underset{d\in\mathcal{D}}{\arg\max} \ 
Q_{\tau}\big\{ \Tstardx \big\}.
\end{equation}
As an example, the above optimal IDR with $\tau=0.5$
will lead to the maximal median of the potential survival time
if each individual in the population follows the treatment recommended.
In some other applications (e.g., distributing aid in a social welfare program), we may target improving the 
lower quantile by considering a smaller value of $\tau$. For an arbitrary monotone transformation $h(\cdot)$, it is known that $h\big( Q_{\tau}\{h(\Tstard) \} \big) = Q_{\tau}\{h(\Tstard)\}$ \citep{koenker2005quantile}. Hence, the same decision will be reached no matter 
the analysis is based on the original survival time or its monotonic transformation.

For complete data, 
quantile criterion was studied in \cite{wang2017quantile} but their work is 
not applicable to survival data, which is also theoretically substantially harder 
with the presence of an infinite-dimensional nuisance parameter due to censoring.
\cite{wahed2009estimation} studied estimating the survival quantiles in 
two-stage randomization designs with fixed IDRs but had not 
investigated the more challenging problem of optimal IDR estimation.

In practice, $\mathcal{D}$ is usually chosen to be  a class of IDRs indexed by a Euclidean parameter for interpretability. Same as \cite{zhang2012robust} and others, we focus on the class of index 
rules
$\mathcal{D}=
\Big\lbrace 
\dbetax = \mbox{I}\big( \bbeta^{{T}}\vX>0\big):\ 
\ |\beta_{1}| = 1,\  \widetilde{\bbeta} \in \mathbb{B} 
\Big\rbrace,
$
where 
$\bbeta = (\beta_1, ...,\beta_p )^T=(\beta_1, \widetilde{\bm{\beta}}^T )^T$,
$\mathbb{B}$ is a compact subset of $ \mathbb{R}^{p-1} $ and $\mbox{I}(\cdot)$
denotes the indicator function. 
For identifiability, we assume there exists a continuous covariate whose coefficient has absolute value one. Without loss of generality, we assume $|\beta_1| = 1$. 
The population parameter 
$\bbeta_0=(\beta_{01}, ...,\beta_{0p} )^T$ indexing the optimal IDR is
\begin{equation*}
\bbeta_0 =\underset{\bbeta\in\mathbb{B}^o}{\arg\max} Q_{\tau}\{\Tstardx\},
\end{equation*}
where $\mathbb{B}^o = \{\bbeta \in \mathbb{R}^p: |\beta_{1}| = 1,\ 
\tilde{\bbeta} \in \mathbb{B} \}$.


\subsection{A robust estimation procedure}\label{sec:estimation procedure}
Based on the observations $\big\{\vX_i,A_i,Y_i,\Delta_i \big\}$, $i=1, \ldots, n$, our goal is to estimate the population parameter 
$\bbeta_0$ indexing the optimal IDR.
It is known that a misspecified generative regression model can result in
severe bias in estimating the optimal treatment \citep{qian2011, zhang2012robust, zhao2012estimating, zhao2015doubly}. We introduce a robust estimator that accounts for censoring while at the same time circumvents the difficulty of specifying a reliable generative regression model.

Given an IDR $d_{\bbeta}(\vX)$, the treatment it would recommend to 
subject $i$ may or may not coincide with the treatment the subject actually received.
Even if $A_i = d_{\bbeta}(\vX_i)$, we may not 
observe $\Tstardbetai$ if the subject is censored.
To obtain a consistent estimator for $ Q_{\tau}\{ \Tstardbetai \}$,
we adapt the induced missing data framework in \cite{zhang2012robust} to time-to-event data. 
Specifically, we consider an artificial missing data structure with the missing data indicator 
$
\Rbetai = \big[A_{i} \dbetaxi +
\big(1-A_{i}\big)\big\lbrace 1- \dbetaxi \big\rbrace \big]\Delta_{i}.
$
The observed outcome $Y_i$ is equal to the potential outcome $\Tstardbetai$ only if  $\Rbetai = 1$.
In this framework, the ``full data'' that we may not completely observe consist of 
$\{\vX_{i}, \Tstardbetai\}_{i=1}^n$, and the observed
data consist of $
\big\{\vX_{i}, \Rbetai,     \Rbetai \Tstardbetai \big\}  = \big\{\vX_{i}, \Rbetai, \Rbetai Y_{i} \big\}_{i=1}^n$.

Let $\pi_{A}(\vX_i)=P(A_i=1|\vX_i)$ be the propensity score; and let  
$ \Gtrue \big(t|\vX, A \big) =  P\big(C > t | \vX, A\big) $ denote the conditional survival function of $ C$ given  $\{\vX, A\}$. Let $\pi_{i}\big(\bbeta\big)	= {P} \bigl\{ \Rbetai  = 1 \mid \vX_{i},\TstarOnei, \TstarZeroi \bigr\}$
be the probability of missingness conditional on the full data. 
We observe 
\begin{align*}
&\pi_{i}\big(\bbeta\big) &	\\
=\  & \pi_{A}(\vX_i)\dbetaxi {P} \bigl[ \Delta_{i}=1 \mid \vX_{i},\TstarOnei, \TstarZeroi, A_{i} =1 \bigr]  & \\
& +(1-\pi_{A}(\vX_i))(1-\dbetaxi) {P} \bigl[ \Delta_{i}=1 \mid \vX_{i},\TstarOnei, \TstarZeroi, A_{i} =0 \bigr] & \\
=\  & \big\{\pi_{A}(\vX_i)\dbetaxi+(1-\pi_{A}(\vX_i))(1-\dbetaxi)\big\} 
\Gtrue \Big( \Tstardbetai \big| \vX, A_i=\dbetaxi \Big). & 
\end{align*}
Note that for the complete cases (corresponding to $ \Rbetai = 1$), we have $Y_i=\Tstardbetai$
and the corresponding
\begin{equation*}
\pi_{i}\big(\bbeta\big)= \
\big[ \pi_{A}(\vX_i)\dbetaxi+ \big(1-\pi_{A}(\vX_i)\big) \big(1-\dbetaxi \big) \big]
\Gtrue \big( Y_i \big| \vX, A_i \big).
\end{equation*}

To estimate $\bbeta_0$, we propose the following two-step estimator. First, we estimate $\Qtaubetax $ by the
following inverse probability weighted estimator
\begin{equation}\label{eq:Qhat}
\QHattaubetax =
\arg\min\limits _{\substack{b}}
\sum_{i=1}^{n}
\frac{\ R_{i}\big( \bbeta \big)}{\widehat{\pi}_i}   \rho_{\tau}\big(Y_{i} - b\big),
\end{equation}	
where $\widehat{\pi}_i$ is an estimate of $\pi_i(\bbeta)$ and  $ \rho_{\tau}\big(u\big) = u\{\tau - I\big(u < 0\big)\}$ is the quantile loss function \citep{koenker2005quantile}. By convention, we define $0/0=0$. Next, employing the policy-search idea, we estimate $\bbeta_0$ by 
\begin{equation}
\label{eq: main_regime_estimator_full}
\widehat{\bbeta}_n = \
\underset{\bbeta  \in \mathbb{B}^o}{\arg\max} \QHattaubetax.
\end{equation}
The above estimator can be computed using the genetic algorithm in R package \texttt{rgenoud} \citep{Mebane2011}.
The estimate of the optimal IDR is $d_{\widehat{\bbeta}_n}\big(\vX\big) = I\big(\widehat{\bbeta}_n^T\vX > 0\big)$.\\

\noindent \textit{Remark 1. }
A key quantity in estimating  $\pi_{i}\big(\bbeta\big)$ is the conditional survival function of the censoring variable $\Gtrue \big(t|\vX, A \big) = P\big(C>t | \vX, A \big)$. There are several approach for estimating $\Gtrue \big(t|\vX, A \big)$. 
For clarity of presentation, as in \cite{goldberg2012q} and \cite{jiang2017estimation}, we assume that 
$C \perp \{\TstarZero, \TstarOne, A, \vX\}$ in the theoretical development. 
This is often satisfied in real applications where 
administrative censoring occurs. In this case, 
$\Gtrue \big(t|\vX, A \big)$ can be estimated by $\widehat{G}_C(\cdot)$, the classical Kaplan-Meier estimator applied to $\{ (Y_i, 1-\Delta_i), i=1,2,...,n\}$. When necessary, we can relax the independent censoring assumption to the conditionally independent censoring assumption $C\perp \{\TstarZero, \TstarOne\} \big| \{\vX, A\}$ and employs the local Kaplan-Meier estimator. Without loss of generality, we assume that the first $n_1$ subjects receive treatment
$A=0$, and the other $(n-n_1)$ subjects receive treatment $A = 1$. The local Kaplan-Meier estimator 
\cite{gonzalez1994asymptotic} for $\Gtrue \big(\cdot|\vX, A=0 \big)$ is given by 
\begin{equation}\label{eq: local_KM_A=0}
\widehat{G}_C\big(\cdot\mid \vX, A=0\big)=\prod_{j=1}^{n_{1}}\Big\{ 1-\dfrac{B_{n_1j}\big( \vX \big)}{\sum_{k=1}^{n_{1}}I\big(C_{k}\geq C_{j}\big)B_{n_1k}\big( \vX \big)}\Big\} ^{\eta_{j}(t)},
\end{equation}
where $\eta_{j}(t)=I\big(C_{j}\leq t,\Delta_{j}=0\big)$,
and $\big\{ B_{n_1k}\big( \vX \big),k=1,...,n_{1}\big\} $
is a sequence of non-negative weights adding up to 1. A popular choice is the
Nadaraya-Watson's type weights for univariate covariate:
$
B_{nk}\big(X\big)=
\big[\sum_{i=1}^{n}K\big(h_{n}^{-1}(X-X_{i})\big)\big]^{-1}K\big(h_{n}^{-1}(X-X_{i})\big)
$,
where $K(\cdot)$ is a positive kernel function  and $h_{n}$ is a sequence of bandwidths converging to zero as $n\rightarrow\infty$. 
We can obtain $\widehat{G}_C\big(\cdot\mid \vX, A=1\big)$ similarly.
A third approach is to estimate the conditional survival function using a working model, such as the Cox proportional hazards regression model, as investigated in \cite{zhao2015doubly}. \\

\noindent \textit{Remark 2. } The Kaplan-Meier estimator $\Ghat$ in 
(\ref{eq:Qhat}) is sometimes unstable at the tail of the distribution.
Practically, a simple approach to improve the stability is by employing an artificial censoring technique in  \cite{Zhou2006}, based on the intuition that any alteration of a random variable's distribution beyond the quantile of interest would
have no impact on the quantile.
Specifically, assume there exists a large positive constant $ M \in \mathbb{R}$ such that
$\sup_{\bbeta}\Qtaub < M$ and $\sup\{t: G_{C}(t)>0 \} >M$. 
The first requirement means the largest achievable $ \tau $th quantile using IDRs belonging to $ \mathcal{D} $ is smaller than $M$; and the second one ensures every data point has a positive probability of not being censored. Note that these conditions are weak, especially if we are interested in lower quantiles.
Let $Y^M = Y\land M$ and  $\Delta^M = \Delta + \big(1-\Delta\big)I\big(Y\geq M\big)$.
Then it is straightforward to show that $\widehat{\bbeta}^M_n $ obtained using the transformed data set $ \big(\vX_i, A_i, Y^M_i,\Delta^M_i\big) $, $ i=1,\ldots,n$, is the same as $\widehat{\bbeta}_n  $ in (\ref{eq: main_regime_estimator_full}).\\

\noindent \textit{Remark 3.} As the optimization problem is nonconcave and nonsmooth,
multiple local optimal may exist. Popular algorithms based on derivatives do not work for
this challenging setting. At the same time, it is impractical to exhaustively enumerate 
all possible solutions and pick the best one.
In our numerical experiments,
we utilize the genetic algorithm in the R package \texttt{rgenoud}, 
which is useful in such a challenging setting when the objective function is
nonconcave and the derivatives do not exist. 
The genetic algorithm (a type of evolutionary algorithm) is inspired from the biological evolution process.
In a genetic algorithm, the problem is encoded in a series of bit strings that are manipulated by the algorithm.
It is a stochastic, population-based algorithm that searches randomly by mutation and crossover among population members.
It is based on searching for the best solutions using inheritance and strengthening of useful features of multiple objects 
of a specific application in the process of imitation of their evolution.
We refer to \cite{mitchell1998} and \cite{Mebane2011} for more detailed description of the algorithm and other
references. In our numerical experience, the algorithm
provides high-quality solutions with desirable statistical properties.

\section{Asymptotic theory}
\label{sec: statistical properties of single-stage estimator}
In this section, we present two results regarding the asymptotic properties of the estimated optimal IDR with survival data.
\begin{itemize}
	\item First, we show that the estimated parameter indexing the optimal IDR has nonstandard asymptotics, which is characterized by the cube-root convergence rate and the non-normal limiting distribution.
	\item Second, we show that under rather weak conditions, which do not require the optimal IDR to be unique at the population or sample level, the
	theoretically optimal value can be estimated at a near $n^{-1/2}$-rate.
	
\end{itemize}
Both results are novel for optimal IDR estimation with time-to-event outcomes. 
The first result corresponds to a nonstandard estimation problem with both finite-dimensional and infinite-dimensional estimation problem. The second result deals with the challenging setting of exceptional law where the optimal IDR is nonunique.

\subsection{Asymptotic distribution of the estimated parameter indexing the optimal IDR}\label{sec31}
Write $\vX = (X_1, ..., X_p)^T \equiv (X_1, \tilde{X}^T)^T$. Let $\Gtrue(\cdot)$ denote the survival function of $C$. Let $ F_{T^{\ast}(0)}(t|\vX)$ and $F_{T^{\ast}(1)}(t|\vX)$ be the cumulative distribution functions of the potential survival times 
$T^{\ast}\big(0\big)$ and $T^{\ast}\big(1\big)$, respectively; and let
$ f_0(t \vert \vX) $ and  $ f_1(t \vert \vX) $ be the corresponding conditional density functions. 
Given any $\bbeta \in \mathbb{B}^p$, let $ f_{\Tstardbeta}(\cdot) $  denote the marginal density function of the distribution of the potential survival time $T^{\ast}(\dbetax)$.

To avoid complications irrelevant to the main results of the paper, we consider data collected from a randomized study where $\pi_{A}(\vX_i)=0.5$, but the results can be extended to observational data under mild assumptions. 
The estimator $\widehat{Q}_{\tau}\big(\Tstardbeta\big)$ in (\ref{eq:Qhat}) simplifies to
\begin{equation}\label{eq:QtauEstimator_Ghat2}
\widehat{Q}_{\tau}\big(\bbeta; \Ghat\big) =
\arg\min\limits _{\substack{b}}
\sum_{i=1}^{n}
\frac{\ R_{i}\big( \bbeta \big)\  \rho_{\tau}\big(Y_{i} - b\big) }
{0.5 \Ghat\big(Y_i\big)},
\end{equation}	
where $\Ghat(\cdot)$ is the classical Kaplan-Meier estimator of $\Gtrue(\cdot)$. 
We then estimate $\bbeta_0$ by 
$\widehat{\bbeta}_n = \underset{\bbeta  \in \mathbb{B}^o}{\arg\max} \widehat{Q}_{\tau}\big(\bbeta; \Ghat\big)$.
Write $\widehat{\bbeta}_n = (\widehat{\beta}_{n1}, \widetilde{\bbeta}_n^T)^T$, where $\widehat{\beta}_{n1}$ satisfies the identifiability condition $|\widehat{\beta}_{n1}| = 1.$ In the proof of Theorem \ref{thm: asymptotic_dis_of_beta_n} in the online supplement, it was shown that $\widehat{\bbeta}_n$ is consistent for $\bbeta_0$. Hence, we have $\widehat{\beta}_{n1} = \beta_{01}$ with probability approaching one. Theorem \ref{thm: asymptotic_dis_of_beta_n} below states the nonstandard convergence rate and non-normal limiting distribution of $\widehat{\bbeta}_n$.

\begin{theorem}\label{thm: asymptotic_dis_of_beta_n}
	Suppose conditions (C1)-(C4) are satisfied. Then as $n\rightarrow \infty$,  
	
	\begin{equation}
	n^{1/3} \big(\widetilde{\bbeta}_n - \widetilde{\bbeta}_{0}\big)  \rightarrow \arg\max\limits _{\substack{t}} \big\{ \Psi(t) + \mathbb{W}(t)  \big\}
	\end{equation}  
	in distribution, where $ \Psi (t) $ is a deterministic function whose form is given in (17) of the online supplement  and  $ \mathbb{W}(t) $ is a mean-zero Gaussian process with covariance function given in (19) in the online supplement.
\end{theorem}

The proof of Theorem \ref{thm: asymptotic_dis_of_beta_n} is given in the online supplement. Theoretical analysis of the asymptotic distribution of 
$\widehat{\bbeta}_n$ in  (\ref{eq:QtauEstimator_Ghat2}) is challenging, as it is defined via a bilevel optimization problem. The proof involves reformulating $\widehat{\bbeta}_n$ as an $M$-estimator with a nonsmooth objective function that has two nuisance parameters: a finite dimensional nuisance parameter $m_0$ and an infinite dimensional nuisance parameter $\Gtrue(\cdot)$. The nonstandard asymptotics arise from the so-called \textit{sharp-edge effect}, see 
\cite{KimPollard90} for an informative example of the shorth estimator that illustrates this phenomenon. It is worth noting that the theory in \cite{KimPollard90} 
can only handle a finite dimensional nuisance parameter, hence is not applicable in our setting. \\

\noindent \textit{Remark 4. }
Our results are related to recent work on non-standard
estimation problem in \citet{banerjee2007}, \cite{sen2010inconsistency}, \cite{matsouaka2014evaluating}, \cite{wang2017quantile}, \cite{shi2018massive}, \cite{patra2018consistent} and \cite{banerjee2019divide}. 
However, none of the above work involves an infinite-dimensional nuisance parameter as we face in the current setting. In fact, our estimation method involves both finite-dimensional and infinite-dimensional nuisance parameters, the role of the latter is for estimating the censoring distribution. Due to the nature of nonstandard asymptotics, the problem is different from and much harder than regular M-estimation problem with 
infinite-dimensional nuisance parameters. Advanced empirical process techniques from \cite{van1996}, \cite{kosorok2008} and \cite{delsol2015semiparametric} were adapted here to help deal with the theoretical challenges.\\

\subsection{Estimating the optimal value with possibly non-unique optimal IDR}\label{subsec_3-optimalvaluetheory}
Besides the optimal IDR itself, a quantity of interest is the optimal value, defined as
\begin{equation}\label{define_Vopt}
V_{opt} = \underset{\bbeta  \in \mathbb{B}^o}{\sup}\ \Qtaubetax.
\end{equation}
This quantity is the maximally achievable marginal $\tau$-th quantile of the potential distribution of all IDRs in the given class of candidate rules. It is an important
measure of the performance of the optimal IDR. A natural estimator of this quantity is $\widehat{V}_n = \widehat{Q}_{\tau} \big(\widehat{\bbeta}_n; \widehat{G}_C \big)$.

Theorem \ref{thm: limit_V} shows that under rather weak conditions, $\widehat{V}_n$ provides a near $n^{-1/2}$-rate estimator for $V_{opt}$.
To avoid the requirement of a unique optimal IDR, 
the derivation of this result is based on recognizing that the following equivalent expressions:
\begin{equation}\label{Vopt_equivalent}
V_{opt} = \sup \Big\{v : \underset{\bbeta  \in \mathbb{B}^o}{\sup} P g\big(\cdot, \bbeta, v, \Gtrue\big) \geq 1-\tau \Big\},
\end{equation}
\begin{equation}
\widehat{V}_{n} = \sup \Big\{v : \underset{\bbeta  \in \mathbb{B}^o}{\sup} P_n g\big(\cdot, \bbeta, v, \Ghat \big) \geq 1-\tau \Big\},
\end{equation}
where
\begin{equation}\label{shark}
g\big(\cdot, \bbeta, v, \Gtrue\big) = \frac{R(\bbeta)}{0.5 \Gtrue(Y)} I(Y-v >0).
\end{equation}
The function $g\big(\cdot, \bbeta, v, \Gtrue\big)$ is motivated by the first-order optimization condition of the estimator $\widehat{Q}_{\tau} \big(\bbeta; \widehat{G}_C \big)$.
A careful inspection reveals that (\ref{define_Vopt}) and (\ref{Vopt_equivalent}) are equivalent.
To see this, we observe that for a given $\bbeta$, $g\big(\cdot, \bbeta, v, G\big)$ is 
a monotonically decreasing function of  $v$, and $Pg\big(\cdot, \bbeta, v, G\big) \geq 1-\tau$
when $v \leq \Qtaub$. Correspondingly, $\bbeta_0$ is the parameter indexing the IDR that achieves this $V_{opt}$.
\begin{theorem}\label{thm: limit_V}
	Suppose conditions (C1) is satisfied. We have 
	$
	\widehat{V}_{n} = V_{opt} + o_p(n^{-1/2 + \gamma_0}),
	$
	for an arbitrary  $\gamma_0 > 0$.
\end{theorem}

\noindent \textit{Remark 5.} It is worth emphasizing that Theorem \ref{thm: limit_V} requires much weaker conditions
than Theorem~\ref{thm: asymptotic_dis_of_beta_n} does. In particular, it does not require the optimal IDR to be unique at the population or sample level. This corresponds to a well known challenging situation where there exists a subpopulation who responds similarly to the two treatment options. If one is willing to assume unique optimal 
IDR, then the above result can be strengthened to parametric convergence rate, i.e., ${n}^{-1/2}$ rate.

\subsection{Smoothed resampling inference}\label{smoothb}
Statistical inference for $\widetilde{\bbeta}_{0}$ is challenging due to the nonstandard asymptotic distribution. A natural idea is to use bootstrap. However, 
the standard nonparametric bootstrap procedure is generally inconsistent for cube-root $M$-estimators (e.g., \cite{Abrevaya2005}, \cite{leger2006bootstrap}) even for the relatively
simpler setting without nuisance functions.
As a remedy, $m$-out-of-$n$ bootstrap \citep{bickel2012resampling},
which draws subsamples of size $m$ from the original sample of size $n$ with replacement, 
has been shown to be consistent for $M$-estimators
with a cube root convergence rate in some settings \citep{delgado2001subsampling}. 
Theoretically, $m$ depends on $n$, tends to infinity with $n$, and
satisfies $m =o(n)$. Practically,
choosing an optimal $m$ is not a simple task. Several data-driven approaches for selecting $m$
were investigated but require intensive computation
(e.g.,\citet{banerjee2007}, \citet{bickel2008choice}, \citet{chakraborty2013},
\citet{qian2021sequential}).


In this subsection, we consider an alternative smoothed resampling-based procedure which is computationally more convenient. This approach is motivated by the alternative expression of $\bbeta_0$
(see the derivation of Lemma 1 in the online supplement),  given by
\begin{equation*}
	\bbeta_0 = \underset{\bbeta \in \mathbb{B}^o}{\arg\max} ~P g(\cdot, \bbeta, V_{opt}, \Gtrue),
\end{equation*}
where $V_{opt}$ is the optimal value and $g\big(\cdot, \bbeta, v, \Gtrue\big)$ is defined in (\ref{shark}). 
That is, $\bbeta_0$ is the parameter indexing the IDR that achieves the optimal
value $V_{opt}$. This naturally leads to an alternative representation of $\widehat{\bbeta}_n$, given by
$
	\widehat{\bbeta}_n = \underset{\bbeta \in \mathbb{B}^o}{\arg\max} ~n^{-1} 
	\sum_{i=1}^n g\left(\cdot,  \bbeta, \widehat{V}_{n}, \Ghat\right).
$

To implement the smoothed resampling-based inference, we first obtain the estimator $\widehat{\bbeta}_n$ and then estimate the optimal value function by
$
\widehat{V}_{n} =
\arg\min\limits _{\substack{b}}
\sum_{i=1}^{n}
\frac{\ R_{i}\big( \widehat{\bbeta}_n \big)\  \rho_{\tau}\big(Y_{i} - b\big) }
{0.5 \Ghat\big(Y_i\big)}.
$
Motivated by \citet{wu2021resampling} for mean-optimal treatment regime with complete data,
we consider the following smoothed estimator 
$$
\overline{\bbeta}_n= \underset{\bbeta \in \mathbb{B}^o}{\arg\max} ~ \frac{1}{n} \sum_{i=1}^{n}
\left(2 A_{i}-1\right)\frac{\Delta_{i} I\left(Y_{i}>\widehat{V}_{n}\right)}{G_{C}\left(Y_{i}\right)} K\left(\frac{\vX_{i}^T \bbeta}{h_n}\right),
$$
where $K(\cdot)$ is a kernel function and $h_n$ is a bandwidth. The kernel function $K(\cdot)$
is only required to satisfy some general conditions, for example, we can take it 
to be the cumulative distribution function of the standard normal distribution. 
Replacing the indicator function in the treatment regime by the kernel function
helps alleviates the sharp edge effect.
Write $\overline{\bbeta}_n = (\overline{\beta}_{n1}, \overline{\widetilde{\bbeta}}_n^T)^T$
Similarly as in \citet{wu2021resampling}, it is expected that 
$\sqrt{nh}\left(\overline{\widetilde{\bbeta}}-\widetilde{\beta}_{0}\right)$ is asymptotically normal. Note that $\overline{\bbeta}_n$ minimizes 
the loss function
$-n^{-1} \sum_{i=1}^{n}
\left(2 A_{i}-1\right)\frac{\Delta_{i} I\left(Y_{i}>\widehat{V}_{n}\right)}{G_{C}\left(Y_{i}\right)} K\left(\frac{\vX_{i}^T \bbeta}{h_n}\right)$,
which is a smoothed estimator of a weighted mis-classification error.
We choose $h_n$ by five-fold cross-validation based on this loss function.

The asymptotic covariance matrix is complex and involves unknown counter-factual distributions. 
For inference, we consider the following perturbed smoothed estimator 
$$
\overline{\bbeta}_n^*= \underset{\bbeta \in \mathbb{B}^o}{\arg\max} ~ \frac{1}{n} \sum_{i=1}^{n}
\xi_i\left(2 A_{i}-1\right)\frac{\Delta_{i} I\left(Y_{i}>\widehat{V}_{n}\right)}{G_{C}\left(Y_{i}\right)} K\left(\frac{\vX_{i}^T \bbeta}{h_n}\right),
$$
where $\xi_{1}, \cdots, \xi_{n}$
are positive random weights independent of the data, with mean one and variance one.
To obtain the bootstrap distribution of $\overline{\bbeta}_n^*$, we repeatedly generate
independent random weights and solve for the smoothed estimator.
Write
$\overline{\bbeta}_n = (\overline{\beta}_{n1}, \overline{\beta}_{n2}, \ldots, 
\overline{\beta}_{np})$, and
$\overline{\bbeta}_n^* = (\overline{\beta}_{n1}^*, \overline{\beta}_{n2}^*, \ldots, 
\overline{\beta}_{np}^*)$,
where $|\overline{\beta}_{n1}^*|=1$.  For $j=2,\ldots,p$, let $\eta_j^{*(\alpha/2)}$ and $\eta_j^{*(1-\alpha/2)}$ be the $(\alpha/2)$-th and $(1-\alpha/2)$-th quantile of the bootstrap distribution of   $(nh_n)^{1/2}(\overline{\beta}_{nj}^*-\overline{\beta}_{nj})$, respectively, where $\alpha$ is a small positive number. We can estimate $\eta_j^{*(\alpha/2)}$ and $\eta_j^{*(1-\alpha/2)}$ from a large number of bootstrap samples. An asymptotic $100(1-\alpha)\%$ bootstrap confidence interval for $\beta_{0j}$,   $j=2,\ldots, p$, is given by
$
\big\{\overline{\beta}_{nj}-(nh_n)^{-1/2}\eta_j^{*(1-\alpha/2)}, \overline{\beta}_{nj}-(nh_n)^{-1/2}\eta_j^{*(\alpha/2)}\big\}.
$.



\section{Estimation of optimal dynamic IDR with censored data.}
In this section, we consider the extension to the dynamic decision problem
which involves multiple decision points. The decision at a later stage can depend on baseline covariates, treatment history, and intermediate variables such as how the subject responds to earlier treatment(s). For survival data, complications arise as the subject may be censored anytime before the end of the study, which results in an incomplete trajectory of treatments.

\subsection{Potential outcome framework}
For clarity, we focus on a two-stage dynamic decision problem with random right censoring.
We consider a setup similar as \cite{jiang2017estimation} but will define the potential outcome more carefully. 
At the beginning of a study (time point 0), baseline covariates $\vX_1$ of patients would be collected, and each of them is assigned one of two first-stage treatment options, say $A_1$ and $A_2$. 
Then the second stage starts from a prespecified time $s$ with $s>0$. 
Additional intermediate covariates $\vX_2$ reflecting the reaction to first stage treatment up to time $s$ would be collected if applicable, and those subjects who remain at risk at time $s$ is assigned one of two first-stage treatment options, say $B_1$ and $B_2$.  
$\{B_1,B_2\}$ may not overlap with $\{A_1,A_2\}$. For example, in making decisions for cancer patients, $\{A_1,A_2\}$ could be induction treatments, while $\{B_1,B_2\}$ may represent maintenance treatment or salvage treatment.


Similarly as in the one-stage setting, the potential survival time is defined when censoring is absent, and we would like to estimate the optimal dynamic IDR with a criterion based on this potential survival time when the real data is complicated by censoring.
Let $D_i$ denote the random treatment at stage $i$ when the subject is eligible. Note that $D_2$ may not exist if the patient is not at risk at time $s$.  
Consider a sequential IDR $\bm{d} = (d_1,d_2)$, where $d_1(\vX_1) \in \{A_1,A_2\}$ and 
$d_2\left( \vX_1, D_1, \vX_2 \right) \in \{B_1,B_2\}$. A subject is considered to be consistent with $\bm{d}$ if he/she receives a first treatment $D_1$ that equals $d_1(\vX_1)$  and a second treatment $D_2$ which equals $d_2(\vX_1, D_1, \vX_2)$  (full compliance); \emph{or} receives treatment $D_1$ complying with the rule $d_1$ at stage one but does not survive long enough to be eligible for stage two treatment. 
Let $ \dot{R}(d_1)= I\Big(\TDOneNo > s\Big) $ indicate the subject's eligibility status for stage two treatment when complying with rule $d_1$, where $\TDOneNo$ is shorthand notation of
$\dot{T}(d_1(\vX_1), \emptyset)$,
which represents the potential survival time if the subject receives $d_1$ without stage-two action. 
Let $\TDOneDtwo$ be the potential survival time if the subject receives the full sequence of treatments $(d_1, d_2)$. Implicitly, $\TDOneDtwo > s$. Let $T^{\ast}(\bm{d})$ be the potential survival time if the subject is consistent with the treatment sequence $\bm{d}$. We can write
\begin{equation}\label{eq: po_dyn}
T^{\ast}(\bm{d}) = \TDOneNo\big(1 - \dot{R}(d_1)\big) + \TDOneDtwo \dot{R}(d_1).
\end{equation}
We are interested in estimating the optimal sequential decision  $\bm{d} = (d_1,d_2)$ in some class $\mathcal{D}$, that is, $\bm{d}_{opt} = \arg\max_{\bm{d}\in\mathcal{D}} Q_{\tau}(\bm{d})$.

Define $H_1 = \{\vX_1 \}$, and define 
$H_2^{\ast}(d_1) = \left\{ \vX_1, d_1, \vX_2^{\ast}(d_1) \right\}$, 
where $\vX_2^{\ast}(d_1)$  denotes the potential intermediate information between decision 1 and decision 2 had the subject started with treatment $d_1(\vX_1)$ and given $\dot{R}(d_1)=1$. 
Denote the set of potential outcomes as
\begin{equation*}
O^{\ast}(\bm{d}) = \left\{\TDOneNo, \ \dot{R}(d_1), \  \dot{R}(d_1)\vX_2^{\ast}(d_1),\  \dot{R}(d_1) d_2\big(H_2^{\ast} (d_1) \big), \   T^{\ast} (\bm{d})\right\}.
\end{equation*}

\subsection{Robust estimation of optimal IDR}
When censoring is absent,
for a given subjects, we would observe the first stage treatment $D_1$ and the corresponding $\dot{R}(D_1)$.
The consistency assumption for 
causal inference, similar as in the one-stage setting, ensures that  the observed survival time $T$ would satisfy
\begin{equation}\label{T_censored_data}
T = \begin{cases}
\sum_{ j\in\{1,2\} } I\{D_1=A_j\} \dot{T}(A_j, \emptyset),   & \mbox{if }  \dot{R} = 0 \\
\sum_{j\in\{1,2\},\ k\in\{1,2\} } I\{D_1 =A_j, D_2 =B_k\} \dot{T}\big(A_j, B_k\big)
& \mbox{if } \dot{R} = 1.
\end{cases}
\end{equation} 

Due to censoring, we may not observe $T$. Denote the actually observed survival time 
under possible censoring as $Y = \min\{T,C\}$, and let $\Delta = I(T \leq C)$ 
be the censoring indicator. Further, let $\Gamma = I(C > s)$ denote whether 
censoring occurs in the first stage. As a result of censoring, 
only those subjects that survived longer than $s$ and are not censored before $s$ 
are eligible for the second-stage treatment, for whom the trajectory observed up to time $s$ is  
$H_{2} = \big\{\vX_{1}, D_1, \vX_{2} \big\}$.
When the trial ends, the observed data is
\begin{equation}\label{observed_data_model}
\left\{ \vX_{i1}, D_1,\ \dot{R}_i,\ \Gamma_i, \ 
\dot{R}_i \Gamma_i \vX_{i2},\ 
\dot{R}_i \Gamma_i D_2,\ Y_i,\ \Delta_i
\right\}, \mbox{ for }  i=1,\ldots,n.
\end{equation}


Based on the observed data, our goal is to estimate the optimal sequential decision
rule within a given class $\mathcal{D}$.  Extending the one-stage formulation, we consider sequential IDR of the form
$ \bm{d}_{\bm{\xi}}=\big\{ d_{1,\bbeta}(H_1) , d_{2,\bm{\zeta}}(H_2)\big\}$, where
$ d_{1,\bbeta}(H_1) = d_{1,\bbeta}(\vX_1) = I \big(\vX_1^T\bbeta > 0\big)$, 
$d_{2,\bm{\zeta}}(H_2) = d_{2,\bm{\zeta}}(\vX_1, D_1, \vX_2) = I \big( H_2^T\bm{\zeta} > 0\big)$ and
$\bm{\xi}=(\bbeta^T, \bm{\zeta}^T)^T$. 
Without loss of generality, we assume that if $d_{1,\bbeta}(\vX_1) = 0$, then the recommended
first-stage treatment is $A_1$, otherwise is $A_2$; and if
$d_{2,\bm{\zeta}}(H_2) = 0$, then the recommended second-stage treatment is $B_1$, otherwise is $B_2$. For identifiability, we assume $\bbeta$ and $\bm{\zeta}$ satisfy  $\vert \beta_1 \vert=1$ and $\vert \zeta_1 \vert=1$. 
Hence $\mathcal{D} = \big\lbrace \bm{d}_{\bm{\xi}} : \bm{\xi}\in \widetilde{\mathbb{C}} \big\rbrace$
where $\widetilde{\mathbb{C}} = 
\{-1,1\}\times \widetilde{\mathbb{B}}\times \{-1,1\}\times \widetilde{\mathbb{Z}} $, 
with $\widetilde{\mathbb{B}}$ being a compact subset of $\mathbb{R}^{p_1 - 1}$, 
$\widetilde{\mathbb{Z}}$ being a compact subset of $\mathbb{R}^{p_2 - 1}$;   
$p_1$ is the dimension of $\vX_1$, and $p_2$ is the dimension of $(\vX_1^T, D_1, \vX_2^T)^T$. The parameter $\bm{\xi}_0$ indexing the optimal dynamic IDR in $\mathcal{D} $ is defined by:
\begin{equation*}
\bm{\xi}_0 = \underset{
	\bm{\xi}\in \widetilde{\mathbb{C}}  }{\arg\max\ } 
Q_{\tau}\big\{T^{\ast}\big( \bm{d}_{\bm{\xi}}\big)\big\},
\end{equation*}
where  $Q_{\tau} \lbrace \cdot \rbrace $ denotes the marginal $\tau$th quantile ($0<\tau<1$),
and $T^{\ast}(\bm{d}_{\bm{\xi}})$ is obtained by setting 
$\bm{d} = \bm{d}_{\bm{\xi}}$ in (\ref{eq: po_dyn}). 
To extend the policy-search method to estimate $\bm{\xi}_0$, we define
\begin{equation*}
\widetilde{R}( \bm{d}_{\bm{\xi}} ) = \Delta 
I\big(D_{1} = d_{1,\bbeta}(\vX_{1})\big) \  
\big[ I (Y \leq s) + I(Y>s) 
I\big(D_{2} = d_{2,\bm{\zeta}}(H_{2})\big) \big].
\end{equation*}
For subjects with $\widetilde{R}( \bm{d}_{\bm{\xi}} ) = 1$, we observe the potential survival 
time of interest, that is, $Y = T^{\ast}(\bm{d}_{\bm{\xi}})$. 

For simplicity in presentation, we assumed that the data are from a sequential multiple 
assignment randomized trial (SMART, \citep{LD00, Murphy08}), where the randomization probabilities at each stage are known by design. That is, at stage one,
$P(D_1 = A_2) = 1- P(D_1 = A_1) = \pi_1$, while at stage two 
$P(D_2 = B_2|Y_i >s, C_i > s) = 1-P(D_2=B_1|Y_i>s, C_i > s) = \pi_2$. 
Given $\bm{d} = (d_1, d_2) \in \mathcal{D}$,
let 
$ \pi_{d_1}(\vX_{i1}) = \pi_1 d_1(\vX_{i1}) + (1-\pi_1)\{1-d_1(\vX_{i1})\} $ denote the probability of compliance to $d_1$ at stage one;
and let
$\pi_{d_2}(H_{i2}) = \pi_2 d_2(H_{i2}) + (1-\pi_2)\{1-d_2(H_{i2})\}$ denote the probability of compliance to $d_2$ at stage two, given that stage one's target potential data is observed and also $Y>s, C>s$.
Overall, the probability to observe $T^{\ast}(\bm{d})$ is  
\begin{equation}\label{eq: lion}
\widetilde{w}_{\bm{d},i} = \
P\left(\widetilde{R}_i(\bm{d})=1 \mid \vX_{i1}, O^{\ast}_i\big(\bm{d}\big) \right) 
= \pi_{d_1}(\vX_{i1})\Gtrue (Y_i) 
\big\{ I(Y_i \leq s) + 
\pi_{d_2}(H_{i2})I (Y_i > s) \big\}.
\end{equation}

We estimate  $ Q_{\tau}\big\{T^{\ast}(\bm{d})\big\} $ by 
\begin{equation*}
\widehat{Q}_{\tau}\big(T^{\ast}(\bm{d}); \Ghat \big)
=
\underset{b}{\arg\min}
\sum_{i=1}^{n}
\frac{\ \widetilde{R}_i( \bm{d})  \rho_{\tau}\big(Y_{i} - b\big) }
{ \widehat{w}_{\bm{d},i}},
\end{equation*}
where $\widehat{w}^{(2)}_{\bm{d},i}$ is obtained by plugging in the
Kaplan-Meier estimator for $G_C$ in (\ref{eq: lion}). 
For brevity, we use the shorthand notation $\widehat{Q}_{\tau}\big(\bm{\xi}; \Ghat \big)$ 
for $\widehat{Q}_{\tau}\big(T^{\ast}(\bm{d}_{\bm{\xi}}); \Ghat \big)$.

Let $L$ denote the end of the study. Assume there exists a constant $\eta > 0$
such that $\Gtrue(L) > \eta > 0$. Furthermore, assume $C$ has a continuously 
differentiable density function which is bounded away from infinity on $(0, L)$.
Also, $m'_0 \triangleq \sup_{\bm{\xi} \in \widetilde{\mathbb{C}}} 
Q_{\tau} \{T^{\ast}(\bm{d}_{\bm{\xi}})\}  < L$. 
Consider an arbitrary treatment sequence $\bm{d} = (d_1 , d_2)$, with 
$d_1(H_1) \in \{A_1, A_2\}$ and $d_2(H_2) \in \{B_1, B_2\}$. Marginally, 
$\TDOneNo$ and $\TDOneDtwo$ have continuous distributions with continuously differentiable density functions.
$\forall \bm{\xi} \in \widetilde{\mathbb{C}}$, 
let $f_{T^{\ast}(\bm{d}_{\bm{\xi}})}(\cdot)$ denote the marginal density function of the distribution of the potential survival time $T^{\ast}(d_{\bm{\xi}})$. 
There exist positive constants $\kappa_1$ and $\delta$, such that 
$\inf_{\bm{\xi} \in \widetilde{\mathbb{C}}} \
\inf_{|m - m'_0| \leq \delta} f_{T^{\ast}(d_{\bm{\xi}})}(m)\geq\kappa_1$.

The following lemma states the consistency of 
$\widehat{Q}_{\tau}\big(\bm{\xi}; \Ghat \big)$
for the marginal quantile of $ T^{\ast}(\bm{d}_{\bm{\xi}})$.
\begin{lemma}\label{lemma: consistency_DTR}
	For all $ d_{\bm{\xi}} \in \mathcal{D} $, we have
	$
	\widehat{Q}_{\tau}\big( \bm{\xi}; \Ghat \big) \rightarrow  Q_{\tau}\{ T^{\ast}(\bm{d}_{\bm{\xi}}) \} 
	$ in probability.
\end{lemma}

Hence, an estimator of the parameter $\bm{\xi}_0$ is
\begin{equation}
\widehat{\bm{\xi}}_n = \underset{
	\bm{\xi}\in \widetilde{\mathbb{C}}  }{\arg\max\ } 
\widehat{Q}_{\tau}\big\{T^{\ast}\big( \bm{d}_{\bm{\xi}}\big); \Ghat \big\}.
\end{equation}
The optimal value function, the maximally achievable marginal $\tau$-th quantile of 
the potential outcome distribution considering all IDRs in $\mathcal{D}$, 
is given by
$
V_{opt} = \underset{
	\bm{\xi}\in \widetilde{\mathbb{C}}  }{\sup\ } 
Q_{\tau}\big\{T^{\ast}(d_{\bm{\xi}})\big\} .
$
An estimator of this quantity is $\widehat{V}_n = \
\widehat{Q}_{\tau}\big( \widehat{\bm{\xi}}_n, \Ghat \big)$.
Similarly as the one-stage setting, we
can re-express $\widehat{V}_n$ as
\begin{equation}\label{eq: Vn_dyn}
\widehat{V}_n = \sup \left\{m: \underset{
	\bm{\xi}\in \widetilde{\mathbb{C}}  }{\sup\ } 
n^{-1} \sum_{i=1}^{n} 
g\left(\cdot, \bm{\xi},m,\Ghat \right)
\right\}
\end{equation}
where
$
g\left(\cdot, \bm{\xi},m,\Gtrue \right) = \frac{\widetilde{R}( \bm{d}) I(Y-m > 0)}{ \widetilde{w}_{\bm{d}}}.
$
The following theorem shows that in the dynamic setting, the optimal value function can be estimated with a near parametric rate.
\begin{theorem}\label{thm: limit_V_dyn}
	For the estimator $\widehat{V}_{n}$ defined in (\ref{eq: Vn_dyn}), we have 
	$
	\widehat{V}_{n} = V_{opt} + o_p(n^{-1/2 + \gamma_0})
	$ 
	for an arbitrary  $\gamma_0 > 0$.
\end{theorem}
It is worth noting that the above result does not require the optimal IDR
to be unique. Similar nonregular asymptotic distribution for 
$\widehat{\bm{\xi}}_n$ can also be established using the same idea as in Section \ref{sec: statistical properties of single-stage estimator} but more complex notations.\\

\noindent \textit{Remark 6.} The method we propose for the dynamic setting is different from the Q-learning approach,
which searches for optimal treatment regimes starting from
the last stage and moving backward.
The Q-learning approach was extended to censored data by \cite{goldberg2012q}. 
There exist several distinct differences between these two methods. First, the proposed method is model-free
in the sense that it does not require to specify an outcome regression model.
The Q-learning approach is model-based and requires a  
survival time model that incorporates both the covariate effects and
treatment-covariate interaction effects. Second,
the proposed method considers a quantile-optimal criterion while 
\cite{goldberg2012q} adopts a restricted mean criterion. Finally,
from a theoretical perspective, this work focuses on the 
statistical properties of the estimated parameters indexing the
optimal IDR while \cite{goldberg2012q} focuses on the finite sample bound
of the generalization error of the estimated optimal IDR.

\section{Numerical studies}
\subsection{Monte Carlo simulations}
We report simulation results for three different settings. In the first example, we estimate one-stage optimal treatment under random censoring; in the second example, we estimate one-stage optimal treatment under covariate-dependent censoring; while in the third example, we consider a two-stage dynamic optimal IDR estimation problem.\\

\noindent{\bf Example 1 (random censoring)}.
We generate the random sample $ \{\vX_{i},A_{i},Y_{i},\Delta_{i}\} $, $ i=1,2,...,n $,  from the model:
$X_{1} \sim U(0,1)$,
$\TstarZero\big| X_1 \sim  \mbox{Weibull}(\mbox{shape} = 1, \mbox{scale} =1) + 1$,   
$\TstarOne \big| X_1 \sim \mbox{Weibull}(\mbox{shape} = 3, \mbox{scale} =0.5 + X_1) + 2X_1$,
$A\big| X,\TstarZero,\TstarOne \sim  \mbox{Bernoulli}(0.5)$.
The response variable in the absence of censoring is generated by $T = \TstarZero\left(1-A\right) + \TstarOne A$.  
The censoring time $C$ has a constant density function  $0.22$ on $(0,2)$ and a constant density function $ 0.07$ on $[2,10)$. 
The observed response is $ Y=  \min\{T, C \}$ and the censoring indicator is $\Delta = I\left\{ T \leq C  \right\}$.
This setup achieves an overall censoring rate of 35\%.

\begin{figure}
	\centering
	\caption{Histograms of $\TstarZero $ and $ \TstarOne $ stratified by $X_1$}
	\includegraphics[width=10cm]{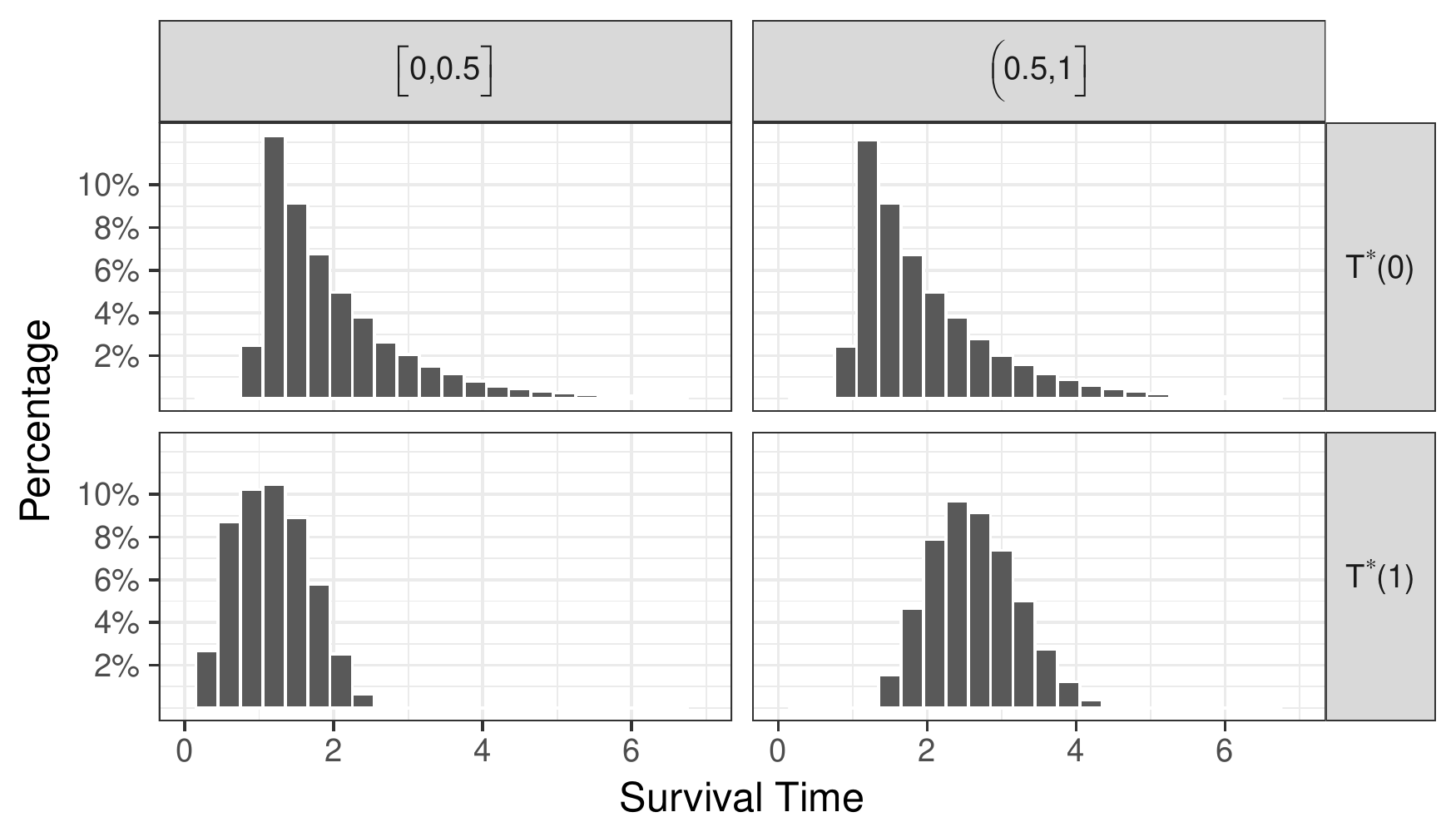} 
	\label{fig1}
\end{figure}

To illustrate the heterogeneous treatment effects,  
we split $ X_1 $ into two strata: $ \left[0, 0.5\right] $ and $ \left(0.5, 1\right]$.
Figure~\ref{fig1} displays the histograms of $\TstarZero$ and $\TstarOne$ in each stratum. 
This plot provides strong evidence that $X_1$ has a qualitative interaction with the treatment.
Intuitively, the optimal IDR should depend on $X_1$. 
We will apply the proposed method to estimate the quantile-optimal IDR for $ \tau=0.25 $ and $\tau=0.5$, respectively.
We consider the following class of  IDRs 
$
\mathcal{D}=\left\{\dbetax=\mbox{I}\left(X_1\beta_1 + \beta_2 > 0\right): \vert\beta_{1}\vert=1, \beta_2 \in \mathbb{R} \right\}.
$
Denote the parameter indexing the $\tau$th quantile optimal IDR in $\mathcal{D}$ by $\bbeta_0^{(\tau)}$.
For each $\tau$, we use a large Monte Carlo data set of size $n = 10^7$ to estimate $\bbeta_0^{(\tau)}$
and the $\tau$th quantile of the potential outcome in the above class of IDRs 
(denoted by $Q_{\tau}$)
and treat the results as population parameter values, see Table~\ref{tab: Example1_truth}. 
Consider, for example, the row corresponding to $\tau=0.5$. We apply the 0.5-quantile optimal IDR to assign treatment in a large independent Monte Carlo sample.
Assume everyone in the population follows the recommended treatment and records his/her outcome.
The median of the potential outcome distribution is 2.258, the first quartile of the 
potential outcome distribution is 1.587. 
%
%


We compare the proposed estimator (denoted by New) with the naive estimator (denoted by Naive), which ignores
censoring and pretends all observations are complete \citep{wang2017quantile}. 
We conduct the simulation experiment with 400 replications
for sample size $ n= $ 300, 500, and 1000. In this experiment, we observed that New always correctly estimates the sign of $\beta_{01}$ for both $\tau=0.25$ and $\tau=0.5$, while Naive has 4\%, 2\%, and 1\% error rate for 
$n=300,500$ and $1000$ respectively in estimating the sign of $\beta_{01}$ for $\tau=0.25$ (0\% error rate 
for $\tau=0.5$). Consideration this phenomenon, we conservatively compare the estimates for $\beta_{02}$
for the two methods in cases where $\hat{\beta}_{01} = 1$. 
Table~\ref{one_cov_RandomCen} summarizes the bias (with standard deviation in the parenthesis) of New and Naive
for estimating $\bbeta_0^{(\tau)}$ and $Q_{\tau}$
for different combinations of $\tau$ and $n$. 
To estimate $Q_{\tau}$, New plugs $\widehat{\bbeta}_n$ into the formula of  $\widehat{Q}_{\tau}\left(\cdot; \Ghat\right)$ in (\ref{eq:QtauEstimator_Ghat2}); and Naive does plugs-in similarly pretending all observation were complete. 
We observe that New has satisfactory performance, while Naive exhibits 
substantial bias for estimating both $\bbeta_0^{(\tau)}$ and $Q_{\tau}$.

\begin{table}
	\caption{ Example 1: Parameters indexing the quantile-optimal IDRs ($\tau=0.25$ and $0.5$) in $\mathcal{D}$ and 
		the $\tau$th quantile of the potential outcome distribution (denoted by $Q_{\tau}$),
		based on a Monte Carlo experiment ($n = 10^7$).}
	\centering 
	\begin{tabular}{r r c c c } \hline 
		$\tau$ & $ {\beta}_{01}^{(\tau)} $ & $ {\beta}_{02}^{(\tau)} $ & $ {Q}_{0.25}$ & $ {Q}_{0.5}$ \\ 
		\hline 
		0.25  & 1 & -0.428 & 1.658 & 2.215 \\
		0.50  & 1 & -0.552 & 1.587 & 2.258 \\
		\hline
	\end{tabular}
\label{tab: Example1_truth}
\end{table}

\begin{table}
	\caption{Bias (with standard deviation in the parenthesis) of New and Naive
		for estimating $\bbeta_0^{(\tau)}$ and $Q_{\tau}$ for Example 1.}
	\centering
	\begin{tabular}{cr cc c cc}
		\hline
		$\tau$ & $ n $   & 	\multicolumn{2}{c}{New}  && \multicolumn{2}{c}{Naive}   \\  
		&& $\beta_{02}^{(\tau)}$ &   $Q_{\tau}$  &&  $\beta_{02}^{(\tau)}$ &
		$Q_{\tau}$ \\ 
		\hline\\
		\multirow{3}{*}{0.25}  	& 300 & 0.005(0.066) & 0.056(0.113) & & -0.025(0.204) & -0.555(0.057)\\ 
		& 500 & -0.001(0.054) & 0.027(0.082)  & & -0.043(0.213) & -0.568(0.050) \\ 
		& 1000 & 0.001(0.043) & 0.020(0.055)  & &  -0.048(0.202) & -0.585(0.031)\\ 
		\hline 
		\multirow{3}{*}{0.50}	& 300 & 0.002(0.098) & 0.048(0.124) & &  0.129(0.082) & -0.618(0.078)\\ 
		& 500 & 0.003(0.080) & 0.023(0.101) & & 0.122(0.064) & -0.617(0.065) \\ 
		& 1000 & -0.002(0.051) & 0.022(0.061) & & 0.134(0.051) & -0.649(0.049)\\
		\hline  
	\end{tabular}
\label{one_cov_RandomCen} 
\end{table}

Finally, we demonstrate the smoothed resampling procedure in Section \ref{smoothb} 
for inference. We consider 90\% and 95\% confidence intervals
for $\beta_{02}$ for $\tau=0.25$ and 0.5, respectively.
The empirical coverage probabilities and average confidence interval lengths
are reported in Table~\ref{inferenceTab} for  
$n=1000$ based on 400 bootstrap samples. 
The observed empirical coverage probabilities are close to the nominal levels with reasonable lengths.

\begin{table}[!htbp]
	\caption{Confidence intervals for $\beta_{02}$ using smoothed resampling}
	\centering
	\begin{tabular}{ccccccccccccccccc}
		\hline
			& &   	\multicolumn{2}{c}{90\% CI}  && \multicolumn{2}{c}{95\% CI}   \\  
		$n$ & $\tau$ &coverage  &length &&coverage  &length \\
		 \hline 
		 500&
		 $\tau =0.5 $ &0.89 & 0.17&& 0.92 &0.21\\
		 &$\tau =0.25 $ &0.91 &0.29&&0.95 &0.34 \\
		 \hline 
		1000&
		 $\tau =0.5 $ &0.88 & 0.13&& 0.93 &0.16\\
		& $\tau =0.25 $ & 0.88 &0.14&& 0.93&0.16
		 \\
		\hline
	\end{tabular}  
	\label{inferenceTab}
\end{table}

\noindent{\bf Example 2 (covariate-dependent censoring)}.
Let $\vX_{i}=\left(X_{i1},1,X_{i2}\right)^{T}$, where
$X_{i1}$, $X_{i2}$ are independent
$\mbox{Uniform}\left(0,1\right)$ random variables.  
The binary treatment $A_i$ is independent of $\vX_{i}$ and satisfies $P(A_i = 1) = 0.5$.
The distribution of censoring variable $C_{i}$ is
\[
C_{i}=\begin{cases}
4+\left(2-X_{i1}\right)\omega_{i}, & \mbox{if }A_{i}=0\\
2+I\left(X_{i1}<0.5 \cup X_{i2}<0.5 \right)+\omega_{i}, & \mbox{if }A_{i}=1 
\end{cases},
\]
where the $\omega_{i}$'s are independent $N(0,1)$ random variables. The survival time $T_i$ is generated by
\begin{equation}
T_{i}=1+X_{i1}+X_{i2}+A_i\left(3-3X_{i1} - 1.5X_{i2}\right)+\left[0.5+A_{i}\left(1+X_{i1}+X_{i2}\right)\right]\epsilon_{i},\label{eq:surv-model}
\end{equation}
where the $\epsilon_{i}$'s are independent normal random variables with mean zero and standard deviation 0.5.
The observed response is $Y_i=\min\left\{T_i, C_i\right\}$. This configuration yields a 30\% censoring rate. 
We consider estimating the $\tau$th quantile optimal IDR ($\tau$=0.1 and 0.25)
within the class $\mathbb{D}=\big\{ \mbox{I}\left(\beta_1 X_1 + \beta_2 + \beta_3 X_2 > 0\right): \vert \beta_1\vert =1, \left(\beta_2, \beta_3\right)^{T}\in {\mathbb{R}^2}\big\}$.
Similarly as for example 1, 
the parameters indexing the quantile-optimal IDRs ($\tau=0.1$ and $0.25$) and the 
the maximally achievable $\tau$th quantile of the potential outcome (denoted by $Q_{\tau}$)	in $\mathbb{D}$
were estimated based on a large Monte Carlo experiment with sample size $n = 10^7$ and treated as population parameter values,
see Table~\ref{tab_Optimal-Treatment-Regimes_MC}.

To incorporate covariate-dependent censoring, we adopt the local Kaplan-Meier estimator 
(with bandwidth $h_n = 0.08, 0.1, 0.12, 0.14$) described in Remark 1 of Section 2
to estimate the propensity score. Table~\ref{tab_Example2_0.1} summarizes 
the simulation results for New and Naive for $n=500$ based on 300 replications.
We observe that New has satisfactory performance and its performance is stable with respect to
different choices of the bandwidth $h$. In contrast, Naive exhibits 
substantial bias for estimating $\bbeta_{02}^{(\tau)}$ and $Q_{\tau}$.\\

\begin{table}
	\caption{
		Parameters indexing the quantile-optimal IDRs ($\tau=0.1$ and $0.25$) and the 
		the maximally achievable $\tau$th quantile of the potential outcome (denoted by $Q_{\tau}$)	in $\mathbb{D}$
		in Example 1,
		based on a Monte Carlo experiment ($n = 10^7$).}
	\centering
	\begin{tabular}{rcccc} 
		\hline
		$\tau$  & $\beta_{01}^{(\tau)}$ & $\beta_{02}^{(\tau)}$ & $\beta_{03}^{(\tau)}$ & $Q_{\tau}({\bm{\beta} })$ \\   
		\hline 
		0.10 & -1 &  0.896  & -0.774   & 1.853 \\ 
		0.25 & -1 &  1.140 & -0.825 &   2.247 \\ 
		\hline 
	\end{tabular} 
\label{tab_Optimal-Treatment-Regimes_MC}	
\end{table}

\begin{table}
	\caption{
		Bias (with standard deviation in the parenthesis) of New and Naive
		for estimating $\bbeta_0^{(\tau)}$ and $Q_{\tau}$ for Example 2.}
	\scalebox{0.8}{
		\centering
		\begin{tabular}{c| cccc ccc}
			\hline 
			&\multicolumn{3}{c}{ $ \tau=0.1$ }&&\multicolumn{3}{c}{$\tau=0.25$ }\\
			Method & $\beta_{02}^{(\tau)}$ & $\beta_{03}^{(\tau)}$  & $Q_{\tau}$
			&&  $\beta_{02}^{(\tau)}$ & $\beta_{03}^{(\tau)}$  & Bias for $Q_{\tau}$ \\ 
			\hline			  
			Naive & -0.107(0.150) & 0.007(0.295) & -0.169(0.065)  && -0.177(0.13) & 0.035(0.215) & -0.206(0.048)\\
			New ($h_n=0.08$)  &  -0.022(0.094) & 0.020(0.162) & 0.025(0.059) && -0.024(0.175) & -0.035(0.265) & -0.029(0.064)  \\ 
			New ($h_n=0.10$) &  -0.016(0.091) & 0.029(0.165) & 0.030(0.058) && 0.019(0.200) & -0.055(0.294) & -0.031(0.067)\\ 
			New ($h_n=0.12$) & -0.009(0.086) & 0.020(0.157) & 0.027(0.058) && -0.014(0.195) & -0.046(0.277) & -0.008(0.065) \\ 
			New ($h_n=0.14$) & -0.020(0.094) & 0.035(0.170) & 0.038(0.054) && 0.002(0.187) & -0.029(0.275) & 0.003(0.069)\\ 
			\hline   
	\end{tabular}}
\label{tab_Example2_0.1}
\end{table}

\noindent{\bf Example 3 (Two-stage dynamic individualized decision rule).}
The simulation setup is motivated by the example in  \citet{jiang2017estimation}. 
The censoring time $ C \sim \mbox{Unif}\left(0, C_0\right) $, where $C_0$ is a positive constant. 
Let $s=1$. Generate the data up to time $s$,  
$\big\lbrace X_1,D_1, \mathsf{Z}^C= \mbox{I}\big(\min\big(T_1, C\big)  >s \big) \big\rbrace,$ 
from the following distributions:
$X_1      \sim \mbox{Unif}\left(0, 4\right)$,
$D_1\mid X_1 \sim \mbox{Bernoulli}(0.5)$ and
$T_1\mid X_1, D_1 \sim \mbox{Exp}\big(\lambda_1 ( X_1, D_1) \big)$
where $\lambda_1(\cdot)$ is a rate function to be specified later, and $\mathsf{Z}^C$ is an auxiliary variable that equals the product of $\dot{R}$ and $\Gamma$ in observed data model (\ref{observed_data_model}). 

If $\mathsf{Z}^C = 1$, then the simulated patient is eligible for stage-two treatment. We 
generate the intermediate covariate $X_2$, second stage treatment $D_2$ and time $T_2$, representing the survival time after time $s$ according to:
$ e \sim \mbox{Unif}\left(0, 2\right)$,
$ X_2\mid X_1, D_1  = 0.5X_1 - 0.4\big(D_1 - 0.5\big) + e$, 
$ D_2\mid X_1, D_1, X_2 \sim \mbox{Bernoulli}(0.5) $,
$ T_2\mid X_1, D_1, X_2, D_2 \sim \mbox{Exp}\big(\lambda_2 (X_1, D_1, X_2, D_2) \big)$,
where $\lambda_2(\cdot)$ is a rate function to be specified later.
For $\mathsf{Z}^C = 1$, the observed survival time and censoring status are $Y = \min\{(s+T_2), C\}$ and $\Delta = \mbox{I}\{(s+ T_2) \leq C\}$, respectively; for $\mathsf{Z}^C = 0$, the observed survival time and censoring status are $Y=\min\{T_1, C\}$ and $\Delta = \mbox{I}\{T_1 \leq C\}$, respectively. 
Let $H_1 = \{X_1\}$ and $H_2 = \{X_1, D_1, X_2\}$.
For rate functions for $T_1$ and $T_2$, consider three scenarios:
\begin{enumerate}
	\item[(a)]      $\lambda_1(H_1, D_1) = 0.5\exp\big(1.75(D_1-0.5)(X_1-2)\big)$,\\
	$\lambda_2(H_2, D_2) = 0.3\exp\big(2.5(D_2-0.4)(X_2-2) - D_1(X_1-2)\big)$
	\item[(b)]      $\lambda_1(H_1, D_1) = 0.2\exp\big(2(D_1-0.5)(-X_1 + 2)\big)$, \\
	$\lambda_2(H_2, D_2) = 0.2\exp\big(1.5 (D_2 - 0.5)(X_2 - 2) + 0.3X_1 + 0.3X_2\big)$
	\item[(c)]     $\lambda_1(H_1, D_1)= 0.3\exp\big(3(D_1-0.3)(X_1-3)\big)$, \\
	$\lambda_2(H_2, D_2)=0.3\exp \big(2(D_2 - 0.5)(X_2-2) - 0.5(D_1-0.3)(X_1 -3)\big)$
\end{enumerate}
For each scenario, we consider two different choice of $C_0$ to achieve 15\% and 40\% overall censoring rate, respectively.

In this setup, $T_1$ serves as the underlying $\sum_{ j\in\{1,2\} } \mbox{I}\{D_1=A_j\} \dot{T}(A_j, \emptyset) $ in equation (\ref{T_censored_data});  $T_2$ is the survival time after initiation of the second stage treatment conditional on $\mathsf{Z}^C=1$, hence $T_2 + s$ serves as $\sum_{j\in\{1,2\},\ k\in\{1,2\} } \mbox{I}\{D_1 =A_j, D_2 =B_k\} \dot{T}\big(A_j, B_k\big)$ in (\ref{T_censored_data}).
By the interaction between  $D_2$ and $H_2$ in $\lambda_2 (\cdot)$ functions, the three scenarios share the same  true optimal second-stage strategy for patients with $\mathsf{Z}^C=1$, which is  $d_{2}^{opt}(H_2) = \mbox{I}(-X_2+2 > 0)$. Suppose the class of IDRs is 
$\mathcal{D} = \big\lbrace 
\bm{d}_{\bm{\xi}} =(d_{1,\bbeta}, d_{2,\bm{\zeta}}): 
d_{1,\bbeta}(X_1) =\mbox{I}\{\beta_{1}X_1 + \beta_2 > 0\}, 
d_{2,\bm{\zeta}}(X_2) =\mbox{I}\{\zeta_{1}X_2 + \zeta_2 > 0\},
\vert \beta_{1} \vert = 1, 
\vert \zeta_{1} \vert = 1
\big\rbrace$.
Thus, $d_{2}^{opt}(H_2)$ is contained in $\mathcal{D} $, and the parameter indexing $d_{2}^{opt}(H_2)$ in $\mathcal{D} $ is   ${\zeta_{01} =-1,\zeta_{02} =2 }$. However, for all three cases, the true value for $\bbeta_0$ indexing the optimal first-stage treatment in $\mathcal{D} $ does not have close form representations. We used grid-search with sample size $n=10^7$ for $\bbeta \in \{-1\}\times [0,4]\cup \{1\}\times [-4,0]$ to obtain $\bbeta_0$ in these cases, where the search space is the largest set of identifiable $\bbeta$ since $X_1$ has support $[0,4]$. 




With the above setup, we generate a random sample $\{X_{i1}, D_{i1}, Z^C_{i} X_{i2}, Z^C_{i}D_{i2}, Y_i, \Delta_i \}$, $i=1,\ldots,n$, where $n=300,500$ and 1000 are considered.
We assume the randomization probability $\pi_1$ and $\pi_2$ are known, and applied the proposed method to estimate the optimal 
IDR with $\tau=0.3$.
Table \ref{DTR_results} reports the simulation estimates of bias and standard deviations of the parameter indexing the quantile optimal dynamic regime $\widehat{\bm{\xi}}_n$. 
It also reports estimates of bias and standard deviation of the plug-in estimator of maximal achievable 0.3-quantile, $\widehat{Q}_{0.3}\{ T^{\ast}(\bm{d}_{\widehat{\bm{\xi}}_n}) \}$.
We observe the proposed method reliably estimated the optimal two-stage IDR. The average biases and standard deviations of $\hat{\beta}_{02}$ and $\hat{\zeta}_{02}$  decrease as sample size increases. The lower censoring rate corresponds to better performance.

{\small
	\begin{table}
		\caption{Simulation Results about the \emph{bias} (standard deviation of the estimates given in the parentheses) of $\widehat{\bm{\xi}}_n$ relative to $\bm{\xi}_0$ and of $\widehat{Q}_{0.3} \big( \widehat{\bm{\xi}}\big)$ relative to ${Q}_{0.3} \big( {\bm{\xi}_0}\big)$. We used 400 replicates. The results for $\hat{\beta}_1$ and $\hat{\zeta}_1$ were omitted, since they ought to be from the set $\{-1, 1\}$ by the normalization method  and our estimator correctly estimate them in each run. True value can be found in Table \ref{DTR_1_truth}.} 
		\centering
		\begin{tabular}{rlllrrr}
			\toprule
			& Case & n & C\% &     \multicolumn{1}{c}{  \mbox{Bias in} ${\beta}_{02}$}  &
			\multicolumn{1}{c}{\mbox{Bias in}   ${\zeta}_{02}$}     &  
			\multicolumn{1}{c}{\mbox{Bias in}   ${Q}_{0.3} \big( {\bm{\xi}_0}\big)$} \\ 
			\hline
			& \multirow{6}{*}{(a)} & 300 & 40 &  -0.052(0.499) & 0.035(0.391) & 0.525(0.445) \\ 
			&  & 500 & 40 &  -0.036(0.416) & 0.031(0.339) & 0.353(0.358) \\
			&  & 1000 & 40 & -0.005(0.317) & 0.043(0.282) & 0.206(0.220) \\     
			\cmidrule(r){3-7}
			&		 & 300 & 15 & 0.016(0.402) & -0.008(0.356) & 0.331(0.345) \\ 
			&  & 500 & 15 & 0.010(0.329) & 0.006(0.303) & 0.247(0.262) \\ 
			&  & 1000 & 15 &  -0.008(0.255) & 0.001(0.230) & 0.139(0.187) \\
			\midrule
			& \multirow{6}{*}{(b)}
			& 300 & 40 & -0.030(0.648) & -0.023(0.371) & 0.280(0.225) \\ 
			&  & 500 & 40 & -0.031(0.581) & -0.015(0.344) & 0.181(0.149) \\  
			&  & 1000 & 40 & -0.016(0.429) & -0.013(0.279) & 0.120(0.105) \\ 
			\cmidrule(r){3-7}
			&    & 300 & 15 & -0.031(0.600) & 0.005(0.332) & 0.235(0.178) \\ 
			&  & 500 & 15 & -0.037(0.485) & 0.012(0.314) & 0.155(0.137) \\
			&  & 1000 & 15 &  0.006(0.414) & -0.004(0.260) & 0.101(0.090) \\ 
			\midrule
			& \multirow{6}{*}{(c)} & 300 & 40 &  -0.084(0.433) & -0.009(0.342) & 0.484(0.396) \\  
			&  & 500 & 40 & -0.056(0.372) & 0.001(0.299) & 0.380(0.347) \\ 
			&  & 1000 & 40 & 0.003(0.277) & 0.008(0.250) & 0.198(0.231) \\
			\cmidrule(r){3-7}
			
			&  & 300 & 15 & -0.074(0.404) & 0.002(0.304) & 0.379(0.329) \\ 
			&  & 500 & 15 & -0.029(0.336) & -0.012(0.254) & 0.261(0.286) \\  
			&  & 1000 & 15 & -0.015(0.249) & 0.001(0.229) & 0.150(0.167) \\  
			
			\bottomrule
		\end{tabular}
	\label{DTR_results} 
	\end{table}
}

\begin{table}
	\caption{True values of $\bm{\xi}_0 = (\bbeta_0^T, \bm{\zeta}_0^T)^T$ indexing the 0.3 quantile-optimal DTR} 
	\centering
	\begin{tabular}{ccccc}
		\toprule
		&   & $\bm{\beta}_0$ &   $\bm{\zeta}_0$ & $Q_{0.3}\big(\bm{\xi}_0\big)$
		\\ 
		\cmidrule(r){3-5}
		& {Case (a)} &  $\big(-1,2.00 \big)^T$ & $\big(-1,2\big)^T$ & 1.524 \\ 
		\cmidrule(r){3-5}
		&  {Case (b)}  &  $\big(1, -1.95\big)^T$   &$\big(-1, 2\big)^T$ & 1.566 \\ 
		\cmidrule(r){3-5}
		&       {Case (c)}       &  $\big(-1,2.94\big)^T$   & $\big(-1, 2\big)^T$ & 2.132\\ 
		\bottomrule
	\end{tabular}
\label{DTR_1_truth}
\end{table}


\newpage
\section{Analysis of GBSG2 study data}
To illustrate the proposed method, we analyze the data from the GBSG2 study conducted by the
German Breast Cancer Study Group \citep{schumacher1994randomized, schmoor1996randomized}.
The study investigated the efficacy of four combinations of treatments: three versus six cycles of chemotherapy with or without the  adjuvant hormonal therapy with Tamoxifen. 
The outcome of interest is the recurrence-free survival time in days.
The dataset, available in the R package \texttt{TH.data} \citep{hothorn2017package},
contains information on 686 patients of whom 56\% had censored outcomes.
The covariates include the age
at diagnosis, the menopausal status, tumour size, tumour grade, the number of positive
lymph nodes, estrogen receptor (ER) and progesterone receptor (PR) expression level
in the tumour tissue. 
Earlier work on this study provided strong evidence that six cycles of chemotherapy is not superior to three cycles with respect to recurrence-free survival. Our analysis therefore focuses on IDRs regarding the assignment of adjuvant Tamoxifen therapy. 
Figure 2 depicts the estimated Kaplan-Meier curves for the groups of patients  with and without Tamoxifen therapy.

\begin{figure}
	\caption{Plot of the Kaplan-Meier estimator of survival functions of $ T $ for $ A=0,1 $ respectively}
	\centering
	\includegraphics[scale=0.5]{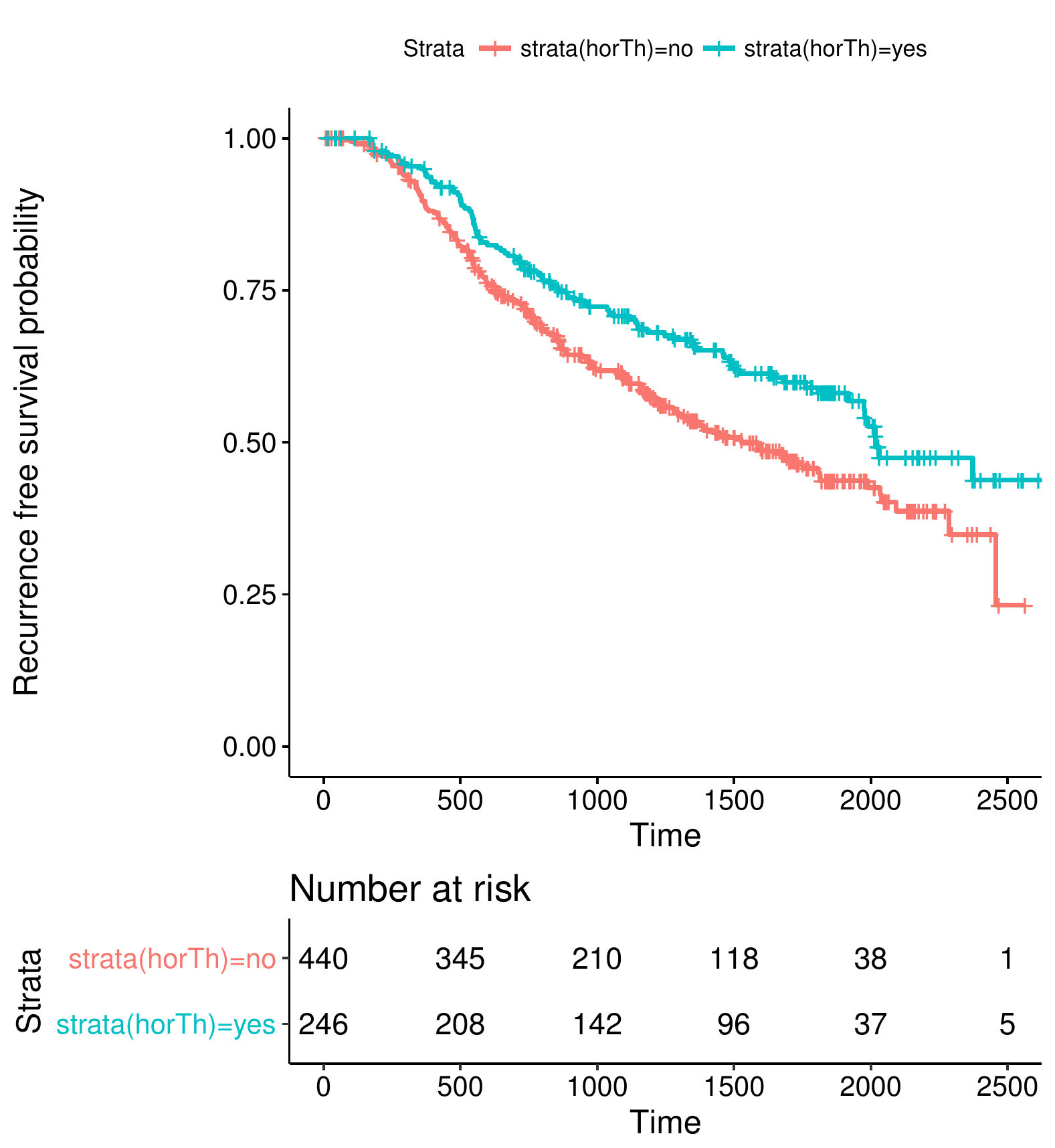}
\end{figure}

First, we estimate the probability of receiving Tamoxifen.  In this study, about two thirds of the recruited patients were randomized, 
and those who were not randomized chose the treatment by personal or professional preference. 
Because the randomization status is masked in the anonymous version of this data, we consider a working propensity score model by logistic regression.
We observe that the randomization status is not significantly associated with the survival \citep{schmoor1996randomized}.
Let $A$ denote the hormonal therapy status ($A=0$: did not receive; $A=1$: received).
We first fit a logistic regression model using $A$ as the response and
all available covariates. We then perform a best subset
selection using the R package \texttt{bestglm} \citep{mcleod2010bestglm}, 
and obtained the model $\mbox{logit}\{\pi_A\left(\vX; \bm{\gamma}\right)\} = \gamma_0 + \gamma_1 \mbox{MNST}$, where $\mbox{MNST}$ is the binary menopausal status of patients. 
Using this selected model, we obtain 
the estimated propensity score 0.203 for premenopausal patients, and 0.472 for postmenopausal patients. 
The dependency of propensity score on $\mbox{MNST}$ is mostly due to a modification on the protocol for randomization starting from the third year of GBSG2 recruitment. 
We also tried the propensity score model with all covariates and found that it leads to almost the same recommendations as measured by the match ratio 
(percentage of times two decision rules make the same treatment recommendations), which is above 98\%
for both of the two classes of regimes under consideration.

Motivated by the extensive work in the medical literature on Tamoxifen's molecular level mechanism 
and its clinical long-term effects, we consider IDRs that depend on the following three variables.
\begin{enumerate}
	\item {ER}: The role of estrogen receptor expression as a predictive factor guiding the allocation of tamoxifen is well
	recognized.
	A large meta-analysis of randomized clinical trials demonstrated that high-ER patients respond better to Tamoxifen compared with low-ER patients \citep{early1998tamoxifen}.
	
	\item {PR}: Progesterone receptor expression is routinely measured for breast cancer patients
	as an important prognostic factor. However, its predictive power for the efficacy of Tamoxifen is still not well understood. 
	It was observed that breast cancer patients with both high ER and high PR (`double positive') have the best chance of surviving \citep{bardou2003progesterone}. 
	
	\item {Age}: Age is an important risk factor in breast cancer. We speculate that it may also contribute to
	 how well patients respond to the adjuvant Tamoxifen therapy.
\end{enumerate}


Because ER and PR are both highly skewed and have the minimal value 0 in this dataset, we 
adopt the transformation $\mbox{LER} = \log_{10}(\mbox{ER}+1)$ and $\mbox{LPR} = \log_{10}(\mbox{PR} + 1)$. 
Age is linearly normalized to be between 0 and 1, and is denoted by {NAGE}.

We first consider the class of IDRs  $\mathcal{D}_1$
that depend on all three variables:
\begin{eqnarray*}
	\mathcal{D}_{1} =\big\{\mbox{I}\big( \beta_1\mbox{LER} + \beta_{2} + \beta_3\mbox{LPR} 
	+ \beta_4\mbox{NAGE} > 0\big):\beta_{1}=1, \beta_2, \beta_3, \beta_4 \in \mathbb{R}\big\}.
\end{eqnarray*}
We restrict the sign of ER to be positive based on evidence from the clinical practice \citep{hammond2010american}.
We estimate the IDR in the class $\mathcal{D}_{1}$ that maximizes the first quartile  ($\tau=0.25$) of the recurrence-free survival time.
We examine the dependence of  the censoring time $C$ on $A$ and all seven prognostic factors by Cox regression and 
conclude that the independent censoring assumption is plausible. 
Furthermore, since GBSG2 has a high censoring rate and relatively short follow-up time, hereafter we use the artificial censoring technique 
(with $M$ being set as 1550 days)in Remark 2 in Section 2.3 to improve stability of the proposed method.

\begin{table}
	\caption{ Estimated parameters indexing the quartile-optimal IDR and 
		90\% $m$-out-of-$n$ bootstrap confidence intervals for the GBSG2 study}
	\centering
	\begin{tabular}{cccc}
		\hline 
		Regimes & $\beta_{2}$ & $\beta_{3}$ & $\beta_{4}$\\
		\hline 
		\multirow{2}{*}{$\mathcal{D}_1$} 
		&-1.23& 0.94 &-0.14\\
		&(-3.39, -0.88)&(0.87, 2.02 )& (-1.21,  2.39 )\\
		\hline 
		\multirow{2}{*}{$\mathcal{D}_2$} &-1.26& 0.97& / \\
		&(-2.41, -1.07)&(0.43, 2.01)& /\\
		\hline 
	\end{tabular}
\label{tab_GBSG2}
\end{table}

The estimated parameter indexing the quartile-optimal IDR is $\widehat{\bbeta}_{n,\mathcal{D}_1}=\left(1,  -1.23,         0.94,     -0.14 \right)^T  $. This regime leads to an estimated 
quartile survival time of  $ \widehat{Q}_{0.25}\left(\widehat{\bbeta}_{n,\mathcal{D}_1}\right)=1246$ days with approximately 81.6\% of patients being recommended to treatment.  In contrast, the  Kaplan-Meier estimator of the 
first quartile of the observed survival time is 727 (90\% confidence interval $=(622, 805)$). 
The first row in Table~\ref{tab_GBSG2} reported the 90\% smoothed bootstrap confidence interval (Section 3.3)
for each coefficient except $\beta_1$ based on 400 bootstrap samples. 
The coefficient of NAGE, $\beta_4$, is insignificant at the 0.1 level, 
which suggests that age may not be an important variable for determining Tamoxifen.

Next, we estimate quartile-optimal IDR in the following simplified class of IDRs to obtain a concise rule,
\begin{equation*}
\mathcal{D}_{2} = \left\{\mbox{I}\left( \beta_1\mbox{LER} + \beta_{2} + \beta_3\mbox{LPR}> 0\right): \right. 
\left. \beta_{1}=1, \beta_2, \beta_3 \in \mathbb{R} \right\}.
\end{equation*}
The estimated parameter indexing the quartile-optimal IDR in $\mathcal{D}_{2}$ is $\widehat{\bbeta}_{n,\mathcal{D}_2}=
\left(   1,   -1.26,  0.97 \right)^T  $, which leads to an estimated 
quartile survival time of  $ \widehat{Q}_{0.25}\left(\widehat{\bbeta}_{n,\mathcal{D}_2}\right)=1246 $ days with
approximately 82.3\% of patients being recommended to treatment. The second row of Table~\ref{tab_GBSG2} 
reported the 90\% smoothed bootstrap confidence intervals for $\beta_2$ and $\beta_3$.
The coefficient for LPR is significant.
Also, because the estimated $\hat{\beta}_3$ for LPR is about 1, we conclude that LPR is as important as the well-established predictive factor LER in developing an IDR that optimizes the first quartile survival time.

\begin{singlespace}
\bibliographystyle{apalike}
\bibliography{censorref}
\end{singlespace}

\section*{Appendix: Regularity Conditions}

We introduce below a set of regularity conditions needed to establish the statistical theory.
The proof of the theory is given in the online supplement.

\begin{description}
	\item[C1] Let $L$ denote the end of the study.
	The censoring variable $C$ has a continuously differentiable density function which is bounded away from infinity on $ \big(0, L\big) $.
	There exists a constant $ \eta>0 $ such that $ \Gtrue(L) > \eta > 0$.  
	The densities $f_0(t \vert \vX) $ and  $ f_1(t \vert \vX) $ are
	uniformly bounded away from infinity, almost surely in $\vX$; 
	and  $\sup_{\bbeta \in \mathbbBo}\Qtaub < L$. There exist positive constants $\kappa_1$ and $\delta$, such that $\inf_{\bbeta \in \mathbbBo} \inf_{|m-m_0|\leq\delta} f_{\Tstardbeta}(m) \geq \kappa_1$, where $m_0 = \sup_{\bbeta \in \mathbbBo} \Qtaub$.
	
	\item[C2] The probability density function of $X_1$ conditional on $\tilde{\vX}$ is continuously differentiable. The angular component of $\vX$, considered as a random element of the sphere $\mathbb{S} \in \mathbb{R}^p$, has a bounded and continuous density.
	
	\item[C3] The population parameter $\bbeta_0 = (\beta_{01}, \widetilde{\bbeta}^T)^T$ indexing the optimal IDR is unique in $\mathbb{B}^o$.
	
	\item[C4] The $(p-1) \times (p-1)$ matrix $\Lambda(\widetilde{ \bbeta}, h)\vert_{\widetilde{ \bbeta} =\widetilde{ \bbeta}_0,h=h_{0}},$	
	defined in the proof of Lemma 1 in the online supplement, is negative definite.
	
\end{description}

\noindent \textit{Remark. }
Condition (C1) is common in survival analysis, where $L$ is the maximum
follow-up time. The survival time is not observed if it exceeds $L$. Condition (C2) has
to do with population parameter identifiability, as discussed in Section \ref{subsection: new_criterion}. We assume
after a possible rearranging of elements in $\vX$, the density of $X_1$ conditional on $\tilde{\vX}$ is
continuously differentiable for every $\tilde{\vX}$  almost surely. If all covariates are discrete, the
problem of estimating an optimal IDR actually becomes simpler in some
sense as there are finite many decision rules. One can directly compare the estimated
value functions. Condition (C3) is standard for index models. Condition (C4) is needed
for evaluating the Hessian matrix when establishing the limiting distribution of 
$\widehat{\bbeta}_{n}$.
\clearpage


\section*{Supplementary material}

# Supplement to "Transformation-Invariant Learning of Optimal Individualized Decision Rules with Time-to-Event Outcomes"

Yu Zhou, Lan Wang, Rui Song and Tuoyi Zhao

## 1 Proof of theory

We first prove Theorem 2, which is useful for studying the properties of $\widetilde{\boldsymbol{\beta}}_n$.

**Proof of Theorem 2**. Write $m_0 = V_{opt}$ and $\widehat{m}_n = \widehat{V}_n$. Recall that as discussed in Section 3.2, motivated by the first-order-optimization condition of the estimator $\widehat{Q}_\tau(\boldsymbol{\beta}, \widehat{G}_C)$, we have the following alternative expression of $m_0$ and $\hat{m}_n$:

$$g(\cdot, \boldsymbol{\beta}, m, G) = \frac{R(\boldsymbol{\beta})}{0.5 G_C(Y)} I(Y - m > 0),$$

$$m_0 = \sup\big\{m : \sup_{\boldsymbol{\beta} \in \mathbb{B}^o} Pg(\cdot, \boldsymbol{\beta}, m, G_C) \geq 1 - \tau\big\},$$

$$\widehat{m}_n = \sup\big\{m : \sup_{\boldsymbol{\beta} \in \mathbb{B}^o} Pg(\cdot, \boldsymbol{\beta}, m, \widehat{G}_C) \geq 1 - \tau\big\}.$$

By the permanence property of Donsker Class (e.g. Example 2.10.8 in Van Der Vaart and

Wellner (1996)), the class of functions

$$\mathcal{F} := \left\{ g(\cdot, \boldsymbol{\beta}, m, G) = \Delta \frac{I\{A = d_{\boldsymbol{\beta}}(\boldsymbol{X})\}}{0.5 G(Y)} I(Y - m > 0) : \boldsymbol{\beta} \in \mathbb{B}^o, m \in [0, L), \right. \quad (1)$$

$$\left. G(\cdot) \text{ is a positive decreasing function uniformly lower bounded by } \eta \text{ over } [0, L) \right\}$$

is *Donsker*, since the set of regimes $d_{\boldsymbol{\beta}}(\cdot)$ and the class of functions $\{I(Y - m > 0) : m \in [0, L)\}$ are both VC-subgraph classes, and $G^{-1}(Y)$ is uniformly bounded in $\mathcal{F}$. Consider the empirical process indexed by functions from $\mathcal{F}$: $\mathbb{G}_n = \sqrt{n}(\mathbb{P}_n - P)$, where $Pg(Z, \boldsymbol{\beta}, m, G) = \int g(z, \boldsymbol{\beta}, m, G) P(dz)$ takes expectation over the distribution of $Z := \{X, A, Y, \Delta\}$; and $\mathbb{P}_n g(Z, \boldsymbol{\beta}, m, G) = n^{-1} \sum_{i=1}^n g(Z_i, \boldsymbol{\beta}_0, m, G)$. We would like to stress that we use the notation $Pg(Z, \boldsymbol{\beta}, m, \widehat{G}_C)$ to mean $Pg(Z, \boldsymbol{\beta}, m, G)$ being evaluated at $G = \widehat{G}_C$ and hence it is a random quantity.

We first consider perturb $m_0$ with a small $\epsilon > 0$. Let $\hat{\varsigma} = (m_0 + \epsilon, \widehat{G}_C)$, $\varsigma_0 = (m_0 + \epsilon, G_C)$. For any given $\boldsymbol{\beta}$ and $G$, the function $g(\cdot, \boldsymbol{\beta}, m, G)$ is decreasing in $m$. Note that $Pg(\cdot, \boldsymbol{\beta}, m, G) = 1 - F_{T^*(d_{\boldsymbol{\beta}})}(m)$, where $F_{T^*(d_{\boldsymbol{\beta}})}(\cdot)$ denotes the marginal cumulative function of the potential survival time $T^*(d_{\boldsymbol{\beta}}(\boldsymbol{X}))$. Observing that $\sup_{\boldsymbol{\beta} \in \mathbb{B}^o} Pg(\cdot, \boldsymbol{\beta}, m_0, G_C) = 1 - \tau$. By Taylor expansion and condition (C1), $\exists \kappa_1 > 0$ such that for all $\epsilon > 0$, $\sup_{\boldsymbol{\beta} \in \mathbb{B}^o} Pg(\cdot, \boldsymbol{\beta}, \varsigma_0) < 1 - \tau - \epsilon \kappa_1$. We have

$$\sup_{\boldsymbol{\beta} \in \mathbb{B}^o} \mathbb{P}_n g(\cdot, \boldsymbol{\beta}, \hat{\varsigma}) < 1 - \tau - \epsilon \kappa_1 + U_{n,1} + \sup_{\boldsymbol{\beta} \in \mathbb{B}^o} \left| (\mathbb{P}_n - P) g(\cdot, \boldsymbol{\beta}, \hat{\varsigma}) \right|, \quad (2)$$

where $U_{n,1} = \sup_{\boldsymbol{\beta} \in \mathbb{B}^o} \left| P\{g(\cdot, \boldsymbol{\beta}, \varsigma_0) - g(\cdot, \boldsymbol{\beta}, \hat{\varsigma})\} \right|$. Note that

$$\begin{aligned}
& P\{g(\cdot, \boldsymbol{\beta}, \varsigma_0) - g(\cdot, \boldsymbol{\beta}, \hat{\varsigma})\} \\
&= P\left\{ 2\Delta \big( A d_{\boldsymbol{\beta}}(\boldsymbol{X}) + (1 - A)(1 - d_{\boldsymbol{\beta}}(\boldsymbol{X})) \big) \big( G_C^{-1}(Y) - \widehat{G}_C^{-1}(Y) \big) I(Y - m_0 - \epsilon > 0) \right\} \\
&\le 2\eta^2 \int \big| \widehat{G}_C(Y) - G_C(Y) \big| P dz
\end{aligned} \quad (3)$$

It follows from Csörgő and Horváth (1983) that $U_{n,1} = o(n^{-1/2 + \gamma_0})$, almost surely, for any $\gamma_0 > 0$.

Since $\mathcal{F}$ defined in (1) is Donsker,

$$\sup_{\boldsymbol{\beta} \in \mathbb{B}^o, m} \left|(\mathbb{P}_n - P)\, g(\cdot, \boldsymbol{\beta}, m, \widehat{G}_C)\right| = O_p(n^{-1/2}). \tag{4}$$

Denote the left hand side of (4) by $U_{n,2}$.

Let the $\epsilon$ in (2) take value $\epsilon_{1,n} = (U_{n,1} + U_{n,2})/\kappa_1$, which is a sequence of positive numbers that converges to zero. Then for all $n$ large enough

$$\sup_{\boldsymbol{\beta}} \mathbb{P}_n g\big(\cdot, \boldsymbol{\beta} \in \mathbb{B}^o, m_0 + \epsilon_{1,n}, \widehat{G}_C\big) < 1 - \tau - \epsilon_{1,n}\kappa_1 + U_{n,1} + U_{n,2} = 1 - \tau.$$

Hence, by the monotonicity of $g$ in $m$, we have $\widehat{m}_n \leq m_0 + \epsilon_{1,n}$ for all $n$ sufficiently large, which implies that $\widehat{m}_n \leq m_0 + o_p(n^{-1/2+\gamma_0})$ for an arbitrarily small $\gamma_0 > 0$.

In the other direction, there exists a positive constant $\kappa_2$ such that $\sup_{\boldsymbol{\beta} \in \mathbb{B}^o} Pg(\cdot, \boldsymbol{\beta}, m_0 - \epsilon, G_C) > 1 - \tau + \epsilon\kappa_2$ for all small enough $\epsilon > 0$. Similarly, we can verify that there exists a sequence of positive numbers $\epsilon_{2,n} = o_p(n^{-1/2+\gamma_0})$ such that $\widehat{m}_n \geq m_0 - \epsilon_{2,n}$ for all $\gamma_0 > 0$. Combining the stochastic lower and upper bounds for $\widehat{m}_n$, we conclude that $\widehat{m}_n = m_0 + o_p(n^{-1/2+\gamma_0})$ for all $\gamma_0 > 0$. $\square$

To derive the limit of $n^{1/3}(\widetilde{\boldsymbol{\beta}}_n - \widetilde{\boldsymbol{\beta}}_0)$ stated in Theorem 1, we next prove a preliminary result Lemma 1, which establishes that $\widetilde{\boldsymbol{\beta}}_n$ converges to $\widetilde{\boldsymbol{\beta}}_0$ at the rate $n^{-1/3}$.

**Lemma 1.** *Suppose conditions (C1)-(C4) are satisfied, then*

$$\|\widetilde{\boldsymbol{\beta}}_n - \widetilde{\boldsymbol{\beta}}_0\| = O_p(n^{1/3}).$$

**Proof of Lemma 1.** We first prove that $\|\widetilde{\boldsymbol{\beta}}_n - \widetilde{\boldsymbol{\beta}}_0\| = o_P(1)$, where $\|\cdot\|$ stands for the Euclidean norm. We note that by the definition of $m_0$ and $\widehat{m}_n$ in the proof of Theorem 2, we obtain an alternative expression of $\boldsymbol{\beta}_0$

$$\boldsymbol{\beta}_0 = \arg\max_{\boldsymbol{\beta} \in \mathbb{B}^o} Pg(\cdot, \boldsymbol{\beta}, m_0, G_C).$$

That is, $\boldsymbol{\beta}_0$ is the parameter indexing the IDR that achieves the optimal value $m_0$. This

naturally leads to an alternative representation of $\widehat{\boldsymbol{\beta}}_n$, given by

$$\widehat{\boldsymbol{\beta}}_n = \arg\max_{\boldsymbol{\beta} \in \mathbb{B}^o} n^{-1} \sum_{i=1}^n g\left(\cdot, \boldsymbol{\beta}, \widehat{m}_n, \widehat{G}_C\right).$$

We will derive the consistency of $\widehat{\boldsymbol{\beta}}_n$ by checking the high-level sufficient conditions (A1)-(A5) in Theorem 1 in Delsol and Van Keilegom (2020), abbreviated as DVK in the sequel. Let $\mathcal{H}$ be the Cartesian product of $\mathbb{R}$ and $\mathcal{G}$, where $\mathcal{G}$ is the class of decreasing functions bounded between $\eta$ and 1. Let $h = (m, G)$ be an arbitrary element in $\mathcal{H}$, let $h_0 = (m_0, G_C)$ denote the true unknown nuisance parameters and let $\widehat{h} = (\widehat{m}_n, \widehat{G}_C)$ be the vector of estimated nuisance parameters. In the following, we will use the supremum metric on $\mathcal{H}$, given by $d_\mathcal{H}(h, h_0) = \max\left\{|m - m_0|, \sup_t |G(t) - G_C(t)|\right\}$. Conditions (A1) and (A2) are trivially satisfied. Condition (A4) is also satisfied for the proposed estimator because the class of functions $\mathcal{F}$ defined in 1 is also *Donsker*. To check (A3) in DVK, we first note that the Kaplan-Meier estimator $\widehat{G}_C$ is uniformly consistent (Csörgő and Horváth (1983)). In addition, Theorem 2 verified that $\widehat{m}_n$ is consistent for $m_0$. Hence, $d_\mathcal{H}(\widehat{h}, h_0) \xrightarrow{P^*} 0$, and (A3) is satisfied. Finally, similarly as in (3), it is straightforward to check $\lim_{d_\mathcal{H}(h, h_0) \to 0} \sup_{\boldsymbol{\beta}} |Pg(\cdot, \boldsymbol{\beta}, h) - Pg(\cdot, \boldsymbol{\beta}, h_0)| = 0$, and (A5) in DVK holds. Thus, all conditions of DVK's Theorem 1 hold. This verifies the consistency of $\widehat{\boldsymbol{\beta}}_n$.

Under the normalization $|\beta_1| = 1$, we know $\widehat{\beta}_{n1} \xrightarrow{a.s.} \beta_{01}$, as $n \to \infty$. In the following, we will derive the convergence rate of $\widetilde{\boldsymbol{\beta}}_n$. For brevity, we only give detailed proof of the convergence rate of $\widetilde{\boldsymbol{\beta}}_n$ under the assumption $\widehat{\beta}_{n1} = \beta_{01} = \beta_1 = 1$, and critical result for the case $\widehat{\beta}_{n1} = \beta_{01} = \beta_1 = -1$ is given in the footnote.

For proving the $n^{-1/3}$ convergence rate of $\widetilde{\boldsymbol{\beta}}_n$, we will verify the high-level conditions (B1)-(B4) of Theorem 2 in DVK (2015). We have already verified $\|\widetilde{\boldsymbol{\beta}}_n - \widetilde{\boldsymbol{\beta}}_0\| = o_P(1)$. By the property of the Kaplan-Meier estimator (Csörgő and Horváth (1983)), we have $d_\mathcal{H}(\widehat{h}, h_0) = o_p(n^{-1/2+\gamma_0})$ for an arbitrarily small $\gamma_0 > 0$. For an arbitrary $\gamma_0 > 0$, set $\nu_n = n^{1/2-\gamma_0}$, then $\nu_n d_\mathcal{H}(\widehat{h}, h_0) = O_{P^*}(1)$. Therefore, Condition (B1) is satisfied. Condition (B4) is fulfilled automatically by the construction of the estimator with $r_n = n^{1/3}$ and $\Phi_n(\delta) = \sqrt{\delta}$. In the following, we will check Conditions (B2) and (B3).

To verify (B2), we consider the following class of functions:

$$\mathcal{F}_{\delta,\delta_1'} = \{g(\cdot,\boldsymbol{\beta},h) - g(\cdot,\boldsymbol{\beta}_0,h) : \|\boldsymbol{\beta} - \boldsymbol{\beta}_0\| \leq \delta, d_{\mathcal{H}}(h,h_0) \leq \delta_1', |\beta_1| = 1, \widetilde{\boldsymbol{\beta}} \in \widetilde{\mathbb{B}}\}.$$

We will first define and demonstrate a square integrable envelope function $M_{\delta,\delta_1'}(Z)$ for the class $F_{\delta,\delta_1'}$.

$$\sup_{\boldsymbol{\beta},h} \{|g(Z,\boldsymbol{\beta},h) - g(Z,\boldsymbol{\beta}_0,h)| : \|\boldsymbol{\beta} - \boldsymbol{\beta}_0\| \leq \delta, d_{\mathcal{H}}(h,h_0) \leq \delta_1'\}$$

$$= \sup_{\boldsymbol{\beta},h} \{2\Delta G^{-1}(Y)A|d_{\boldsymbol{\beta}}(\boldsymbol{X}) - d_{\boldsymbol{\beta}_0}(\boldsymbol{X})|I(y - m < 0) : \|\boldsymbol{\beta} - \boldsymbol{\beta}_0\| \leq \delta, d_{\mathcal{H}}(h,h_0) \leq \delta_1'\}$$

$$\leq \sup_{\boldsymbol{\beta},h} \{2\eta^{-1}I(d_{\boldsymbol{\beta}}(\boldsymbol{X}) \neq d_{\boldsymbol{\beta}_0}(\boldsymbol{X})) : \|\boldsymbol{\beta} - \boldsymbol{\beta}_0\| \leq \delta, d_{\mathcal{H}}(h,h_0) \leq \delta_1'\}$$

$$\leq 2\eta^{-1}\{I[\delta\|\boldsymbol{X}\| > \boldsymbol{X}^T\boldsymbol{\beta}_0 > 0] + I[\delta\|\boldsymbol{X}\| > -\boldsymbol{X}^T\boldsymbol{\beta}_0 \geq 0]\}$$

$$\leq 2\eta^{-1}I\left[\frac{\delta}{\|\boldsymbol{\beta}_0\|} > \frac{|\boldsymbol{X}^T\boldsymbol{\beta}_0|}{\|\boldsymbol{X}\|\|\boldsymbol{\beta}_0\|}\right] =: M_{\delta,\delta_1'}(Z). \tag{5}$$

Clearly, $E[M_{\delta,\delta_1'}^2(Z)] = 4\eta^{-2}P\{\frac{\delta}{\|\boldsymbol{\beta}_0\|} > |\frac{\boldsymbol{X}^T}{\|\boldsymbol{X}\|}\frac{\boldsymbol{\beta}_0}{\|\boldsymbol{\beta}_0\|}|\}$, where we need to evaluate the probability of the dot product between $\frac{\boldsymbol{X}}{\|\boldsymbol{X}\|}$ and $\frac{\boldsymbol{\beta}_0}{\|\boldsymbol{\beta}_0\|}$ being between $(-\frac{\delta}{\|\boldsymbol{\beta}_0\|}, \frac{\delta}{\|\boldsymbol{\beta}_0\|})$. Let $\Phi_n(\delta) = \sqrt{\delta}$. By Condition (C2) in Subsection 3.1 on the density of the angular component of $\boldsymbol{X}$, $\exists\, \delta_0 > 0, \exists\, K_0 > 0$, such that $\forall\, \delta < \delta_0$,

$$E[M_{\delta,\delta_1'}^2(Z)] = 4\eta^{-2}E\left\{I\left[\frac{\delta}{\|\boldsymbol{\beta}_0\|} > \left|\frac{\boldsymbol{X}^T}{\|\boldsymbol{X}\|}\frac{\boldsymbol{\beta}_0}{\|\boldsymbol{\beta}_0\|}\right|\right]\right\} \leq K_0^2\Phi_n^2(\delta). \tag{6}$$

To verify Condition (B2) in DVK, it is sufficient to show there exists a positive constant $K_1$ such that for all $\delta \leq \delta_0$,

$$E\left[\sup_{\|\boldsymbol{\beta}-\boldsymbol{\beta}_0\|<\delta,\, d_{\mathcal{H}}(h,h_0)\leq \delta_1'} |\mathbb{P}_n g(\cdot,\boldsymbol{\beta},h) - \mathbb{P}_n g(\cdot,\boldsymbol{\beta}_0,h) - Pg(\cdot,\boldsymbol{\beta},h) + Pg(\cdot,\boldsymbol{\beta}_0,h)|\right]$$

$$\leq K_1 n^{-1/2}\sqrt{E[M_{\delta,\delta_1'}^2]}, \tag{7}$$

Notice that $\mathcal{F}_{\delta,\delta_1'}$ is a VC-subgraph class. By Theorem 2.6.7 of Van Der Vaart and Wellner

(1996), $\mathcal{F}_{\delta,\delta_1'}$ satisfies the uniform entropy condition (Van Der Vaart and Wellner, 1996) and

$$\sup_{\delta<\delta_0,V} \int_0^1 \sqrt{1+\log N\big(\epsilon\|M_{\delta,\delta_1'}\|_{L_2(V)}, \mathcal{F}_{\delta,\delta_1'}, L_2(V)\big)}\, d\epsilon < \infty, \tag{8}$$

where the supremum on distribution $V$ takes over all discrete probability measure such that $\|M_{\delta,\delta_1'}\|_{L_2(V)} > 0$. By Van Der Vaart and Wellner (1996) page 244, (8) implies (7) and hence Condition (B2) hold.

To verify Condition (B3), it is sufficient to prove the set of conditions in Remark 2(v) of DVK. We need to calculate the gradient and Hessian matrix of $Pg(\cdot, \boldsymbol{\beta}, h)$ with respect to $\widetilde{\boldsymbol{\beta}}$. Note that we assumed $\beta_{01} = 1$ in this main proof (the case $\beta_{01} = -1$ is only slightly different). We have

$$\begin{aligned}
&Pg(\cdot, \boldsymbol{\beta}, h) \\
&= E\{d_{\boldsymbol{\beta}}(\boldsymbol{X})\Delta G^{-1}(Y)I(Y>m)|A=1\} + E\{(1-d_{\boldsymbol{\beta}}(\boldsymbol{X}))\Delta G^{-1}(Y)I(Y>m)|A=0\} \\
&= E_{\boldsymbol{X}}\big[I(\boldsymbol{X}^T\boldsymbol{\beta}>0)R_1(m,G|\boldsymbol{X})\big] + E_{\boldsymbol{X}}\big[I(\boldsymbol{X}^T\boldsymbol{\beta}\leq 0)R_0(m,G|\boldsymbol{X})\big],
\end{aligned}$$

by the iterative expectation formula; where

$$R_1(m,G|\boldsymbol{X}) = E_C\Big[I(C>m)\int_m^C G^{-1}(u)f_1(u|\boldsymbol{X})du\Big|\boldsymbol{X}\Big],$$

$$R_0(m,G|\boldsymbol{X}) = E_C\Big[I(C>m)\int_m^C G^{-1}(u)f_0(u|\boldsymbol{X})du\Big|\boldsymbol{X}\Big],$$

with $f_1(\cdot|\boldsymbol{X})$ being the conditional density function of $T^*(1)$ given $\boldsymbol{X}$ and $f_0(\cdot|\boldsymbol{X})$ being the conditional density function of $T^*(0)$ given $\boldsymbol{X}$. Define $q(h,\boldsymbol{X}) = R_1(m,G|\boldsymbol{X}) - R_0(m,G|\boldsymbol{X})$, then $Pg(\cdot,\boldsymbol{\beta},h) = E_{\boldsymbol{X}}\{R_0(h|\boldsymbol{X}) + q(h,\boldsymbol{X})I(\boldsymbol{X}^T\boldsymbol{\beta}>0)\}$. Under the restriction $\beta_1 = 1$, let $f_{\boldsymbol{X}_1|\widetilde{\boldsymbol{X}}}(t)$ be the conditional density of $\boldsymbol{X}_1$ given $\widetilde{\boldsymbol{X}}$. The gradient of

6

$Pg(\cdot, \boldsymbol{\beta}, h)$ with respect to $\widetilde{\boldsymbol{\beta}}$ is

$$\Gamma(\widetilde{\boldsymbol{\beta}}, h) = \frac{\partial}{\partial \widetilde{\boldsymbol{\beta}}} E_{\boldsymbol{X}}\big[q(h, \boldsymbol{X}) I(\boldsymbol{X}_1 > -\tilde{\boldsymbol{X}}^T \widetilde{\boldsymbol{\beta}})\big]$$

$$= \frac{\partial}{\partial \widetilde{\boldsymbol{\beta}}} E_{\tilde{\boldsymbol{X}}}\Big\{ \int_{-\tilde{\boldsymbol{X}}^T \widetilde{\boldsymbol{\beta}}}^{\infty} f_{\boldsymbol{X}_1|\tilde{\boldsymbol{X}}}(t) q(h, (t, \tilde{\boldsymbol{X}}^T)^T) dt \Big\}$$

$$= E_{\tilde{\boldsymbol{X}}}\big\{ \tilde{\boldsymbol{X}} f_{\boldsymbol{X}_1|\tilde{\boldsymbol{X}}}(-\tilde{\boldsymbol{X}}^T \widetilde{\boldsymbol{\beta}}) q(h, (-\tilde{\boldsymbol{X}}^T \widetilde{\boldsymbol{\beta}}, \tilde{\boldsymbol{X}}^T)^T) \big\}.$$

Hence,
$$\Gamma(\widetilde{\boldsymbol{\beta}}, h)|_{\widetilde{\boldsymbol{\beta}} = \widetilde{\boldsymbol{\beta}}_0} = E_{\tilde{\boldsymbol{X}}}\big\{ \tilde{\boldsymbol{X}} f_{\boldsymbol{X}_1|\tilde{\boldsymbol{X}}}(-\tilde{\boldsymbol{X}}^T \widetilde{\boldsymbol{\beta}}_0) q(h, (-\tilde{\boldsymbol{X}}^T \widetilde{\boldsymbol{\beta}}_0, \tilde{\boldsymbol{X}}^T)^T) \big\}. \tag{9}$$

The Hessian matrix of $Pg(\cdot, \boldsymbol{\beta}, h)$ with respect to $\widetilde{\boldsymbol{\beta}}$ is

$$\Lambda(\widetilde{\boldsymbol{\beta}}, h) = \frac{\partial}{\partial \widetilde{\boldsymbol{\beta}}} E_{\tilde{\boldsymbol{X}}}\Big\{ \tilde{\boldsymbol{X}} f_{\boldsymbol{X}_1|\tilde{\boldsymbol{X}}}(-\tilde{\boldsymbol{X}}^T \widetilde{\boldsymbol{\beta}}) q(h, (-\tilde{\boldsymbol{X}}^T \widetilde{\boldsymbol{\beta}}, \tilde{\boldsymbol{X}}^T)^T) \Big\}$$

$$= E_{\tilde{\boldsymbol{X}}}\Big[ -\tilde{\boldsymbol{X}} \tilde{\boldsymbol{X}}^T \big\{ \frac{\partial}{\partial t} f_{\boldsymbol{X}_1|\tilde{\boldsymbol{X}}}(t)|_{t=-\tilde{\boldsymbol{X}}^T \widetilde{\boldsymbol{\beta}}} \big\} q(h, (-\tilde{\boldsymbol{X}}^T \widetilde{\boldsymbol{\beta}}, \tilde{\boldsymbol{X}}^T)^T)$$

$$- \tilde{\boldsymbol{X}} \tilde{\boldsymbol{X}}^T f_{\boldsymbol{X}_1|\tilde{\boldsymbol{X}}}(-\tilde{\boldsymbol{X}}^T \widetilde{\boldsymbol{\beta}}) \big\{ \frac{\partial}{\partial t} q(h, (t, \tilde{\boldsymbol{X}}^T)^T)|_{t=-\tilde{\boldsymbol{X}}^T \widetilde{\boldsymbol{\beta}}} \big\} \Big].$$

This Hessian matrix evaluated at $h = h_0$ is [1]:

$$\Lambda(\tilde{\boldsymbol{\beta}}, h)|_{\tilde{\boldsymbol{\beta}}=\tilde{\boldsymbol{\beta}}_0, h=h_0}$$
$$= E_{\tilde{\boldsymbol{X}}}\big(-\tilde{\boldsymbol{X}}\tilde{\boldsymbol{X}}^T\big[\{\frac{\partial}{\partial t} f_{\boldsymbol{X}_1|\tilde{\boldsymbol{X}}}(t)|_{t=-\tilde{\boldsymbol{X}}^T\tilde{\boldsymbol{\beta}}}\} q\big(h_0, \big(-\tilde{\boldsymbol{X}}^T\tilde{\boldsymbol{\beta}}_0, \tilde{\boldsymbol{X}}^T\big)^T\big)$$
$$+ f_{\boldsymbol{X}_1|\tilde{\boldsymbol{X}}}\big(-\tilde{\boldsymbol{X}}^T\tilde{\boldsymbol{\beta}}_0\big)\{\frac{\partial}{\partial t} q\big(h_0, (t, \tilde{\boldsymbol{X}}^T)^T\big)\big|_{t=-\tilde{\boldsymbol{X}}^T\tilde{\boldsymbol{\beta}}_0}\}\big])$$
$$= E_{\tilde{\boldsymbol{X}}}\big(-\tilde{\boldsymbol{X}}\tilde{\boldsymbol{X}}^T\big[\{\frac{\partial}{\partial t} f_{\boldsymbol{X}_1|\tilde{\boldsymbol{X}}}(t)|_{t=-\tilde{\boldsymbol{X}}^T\tilde{\boldsymbol{\beta}}}\} \times \big(F_{T^*(0)|\boldsymbol{X}} - F_{T^*(1)|\boldsymbol{X}}\big)[m_0 \mid \boldsymbol{X} = \big(-\tilde{\boldsymbol{X}}^T\tilde{\boldsymbol{\beta}}_0, \tilde{\boldsymbol{X}}^T\big)^T\big]$$
$$+ f_{\boldsymbol{X}_1|\tilde{\boldsymbol{X}}}\big(-\tilde{\boldsymbol{X}}^T\tilde{\boldsymbol{\beta}}_0\big)[\frac{\partial}{\partial t}\big(F_{T^*(0)|\boldsymbol{X}} - F_{T^*(1)|\boldsymbol{X}}\big)\{m_0 \mid \boldsymbol{X} = (t, \tilde{\boldsymbol{X}}^T)^T\}\bigg|_{t=-\tilde{\boldsymbol{X}}^T\tilde{\boldsymbol{\beta}}_0}\big]\big])$$

This $(p-1) \times (p-1)$ matrix is assumed negative definite in Condition (C4).

Furthermore, by (9), the consistency of $\widetilde{\boldsymbol{\beta}}_n$, the property of Kaplan-Meier estimator and the fact that $\Gamma(\widetilde{\boldsymbol{\beta}}_0, h_0) = 0$, it is straightforward to show $\Gamma(\widetilde{\boldsymbol{\beta}}_0, \widehat{h}) = o_p(n^{-1/2+\gamma_0})$ for an arbitrarily small $\gamma_0 > 0$. So $\|\Gamma(\widetilde{\boldsymbol{\beta}}_0, \widehat{h})\| = O_p(n^{-1/3})$. Hence, Condition (B3) of DVK is also satisfied. Summarizing the above, we have proved the lemma.□.

**Proof of Theorem 1**. We first introduce some new notation. Define

$$B(\boldsymbol{\beta}, h) = Pg(\cdot, \boldsymbol{\beta}, h) - Pg(\cdot, \boldsymbol{\beta}_0, h),$$
$$B_n(\boldsymbol{\beta}, h) = \mathbb{P}_n g(\cdot, \boldsymbol{\beta}, h) - \mathbb{P}_n g(\cdot, \boldsymbol{\beta}_0, h).$$

And, for any $\delta > 0$, define a local upper bound function:

$$M_\delta(\,\cdot\,) \geq \sup_{\|\tilde{\boldsymbol{\beta}}-\tilde{\boldsymbol{\beta}}_0\|\leq\delta} |g(\cdot, \boldsymbol{\beta}, h_0) - g(\cdot, \boldsymbol{\beta}_0, h_0)|, \qquad (10)$$

---

[1] In the case where $\beta_{01} = \beta_1 = -1$, the gradient is the same formula except changing $-\tilde{\boldsymbol{X}}^T\tilde{\boldsymbol{\beta}}$ to $\tilde{\boldsymbol{X}}^T\tilde{\boldsymbol{\beta}}$ in every $q(h, \cdot)$; the Hessian matrix of $Pg(\cdot, \boldsymbol{\beta}, h)$ with respect to $\tilde{\boldsymbol{\beta}}$ evaluated at $\tilde{\boldsymbol{\beta}} = \tilde{\boldsymbol{\beta}}_0, h = h_0$ is:

$$\Lambda^*(\tilde{\boldsymbol{\beta}}, h)|_{\tilde{\boldsymbol{\beta}}=\tilde{\boldsymbol{\beta}}_0, h=h_0}$$
$$= E_{\tilde{\boldsymbol{X}}}\{-\tilde{\boldsymbol{X}}\tilde{\boldsymbol{X}}^T(\{\frac{\partial}{\partial t} f_{\boldsymbol{X}_1|\tilde{\boldsymbol{X}}}(t)|_{t=\tilde{\boldsymbol{X}}^T\tilde{\boldsymbol{\beta}}}\} \times \big(F_{T^*(1)|\boldsymbol{X}} - F_{T^*(0)|\boldsymbol{X}}\big)[m_0 \mid \boldsymbol{X} = (\tilde{\boldsymbol{X}}^T\tilde{\boldsymbol{\beta}}_0, \tilde{\boldsymbol{X}}^T)^T]$$
$$+ f_{\boldsymbol{X}_1|\tilde{\boldsymbol{X}}}\big(-\tilde{\boldsymbol{X}}^T\tilde{\boldsymbol{\beta}}_0\big) \times [\frac{\partial}{\partial t}\big(F_{T^*(1)|\boldsymbol{X}} - F_{T^*(0)|\boldsymbol{X}}\big)\{m_0 \mid \boldsymbol{X} = (t, \tilde{\boldsymbol{X}}^T)^T\}\bigg|_{t=\tilde{\boldsymbol{X}}^T\tilde{\boldsymbol{\beta}}_0}])\}.$$

which dominates the class of functions $\mathcal{M}_\delta = \{g(\cdot, \boldsymbol{\beta}, h_0) - g(\cdot, \boldsymbol{\beta}_0, h_0) : \|\widetilde{\boldsymbol{\beta}} - \widetilde{\boldsymbol{\beta}}_0\| \leq \delta, \beta_{01} = \beta_1\}$.

The proof involves verifying Conditions (C1)-(C10) of Theorem 3 in DVK. We first note that (C1) is satisfied by the proof for Lemma 1. Condition (C2) holds naturally in our case with $E = \mathbb{R}^{p-1}$. Condition (C6) is trivially satisfied under our regularity conditions. Condition (C8) is also satisfied since we constructed $\hat{\boldsymbol{\beta}}$ as the maximizer of the random function $\mathbb{P}_n g(\cdot, \boldsymbol{\beta}, \hat{h})$. We will verify the other conditions.

Condition (C3) can be verified using techniques similar as those for proving (B2) in proof of Lemma 1. More specifically, (C3) holds when the following two conditions are satisfied: there exist a function $l$ and a positive constant $\delta_0 < \eta/2$ such that for all $\delta_2, \delta_3 < \delta_0$, and for an arbitrarily small $\gamma_0 > 0$,

$$n^{2/3} l(n^{-1/3}\delta_2, n^{-1/2+\gamma_0}\delta_3) = o(\sqrt{n}), \tag{11}$$

and

$$E\left[\sup_{\substack{\|\widetilde{\boldsymbol{\beta}} - \widetilde{\boldsymbol{\beta}}_0\| \leq n^{-1/3}\delta_2,\ \beta_{01}=\beta_1, \\ d_\mathcal{H}(h, h_0) \leq n^{-1/2+\gamma_0}\delta_3}} \left|B_n(\boldsymbol{\beta}, h) - B(\boldsymbol{\beta}, h) - B_n(\boldsymbol{\beta}, h_0) + B(\boldsymbol{\beta}, h_0)\right|\right]$$
$$\leq n^{-1/2} l(n^{-1/3}\delta_2, n^{-1/2+\gamma_0}\delta_3). \tag{12}$$

To verify the above conditions, let us consider the class:

$$\mathcal{F}_{\delta_2, \delta_3} = \{g(\cdot, \boldsymbol{\beta}, h) - g(\cdot, \boldsymbol{\beta}, h_0) - g(\cdot, \boldsymbol{\beta}_0, h) + g(\cdot, \boldsymbol{\beta}_0, h_0) :$$
$$\|\widetilde{\boldsymbol{\beta}} - \widetilde{\boldsymbol{\beta}}_0\| \leq n^{-1/3}\delta_2,\ \beta_{01} = \beta_1, d_\mathcal{H}(h, h_0) \leq n^{-1/2+\gamma_0}\delta_3\},$$

where $0 < \delta_0 < \eta/2$ and $0 < \delta_2, \delta_3 < \delta_0$.

It is straightforward to deduce that the random function $|g(\cdot, \boldsymbol{\beta}, h) - g(\cdot, \boldsymbol{\beta}, h_0)|$ is upper bounded by a constant $4\eta^{-2} n^{-1/2+\gamma_0}\delta_3$ (e.g., see (3)); as is the case with $|g(\cdot, \boldsymbol{\beta}_0, h) - g(\cdot, \boldsymbol{\beta}_0, h_0)|$. Therefore, an envelope function of the class $\mathcal{F}_{\delta_2, \delta_3}$ is:

$$F_{\delta_2, \delta_3}(Z) := 8\eta^{-2} n^{-1/2+\gamma_0}\delta_3. \tag{13}$$

It is straightforward to prove that $\mathcal{F}_{\delta_2,\delta_3}$ is a VC-class. Then $\mathcal{F}_{\delta_2,\delta_3}$ has finite uniform entropy integral. By Van Der Vaart and Wellner (1996), page 244, the left hand side of (12) is upper bounded by $\sqrt{E[F^2_{\delta_2,\delta_3}(Z)]}/\sqrt{n}$ up to a multiplicative constant. This, coupled with (13), inspired us to set $l(n^{-1/3}\delta_2, n^{-1/2+\gamma_0}\delta_3) = K_2 n^{-1/2+\gamma_0}\delta_3$ for a constant $K_2 > 8$. Then $\sqrt{E[F^2_{\delta_2,\delta_3}(Z)]} \leq l(n^{-1/3}\delta_2, n^{-1/2+\gamma_0}\delta_3)$. Then, for all positive parameters $\delta_2, \delta_3 < \delta_0$, both (11) and (12) hold, implying that (C3) in DVK holds.

Condition (C4) concerns the function $M_\delta(\cdot)$ defined in (10). Observe that the envelope function $M_{\delta,\delta'_1}$ in (5) for the class $\mathcal{F}_{\delta,\delta'_1}$ does not depend on nuisance parameters and is also an envelope function for $\mathcal{M}_\delta$. So we let

$$M_\delta(\cdot) := 2\eta^{-1} I\left[\frac{\delta}{\|\beta_0\|} > \frac{|\boldsymbol{X}^T \beta_0|}{\|\boldsymbol{X}\|\|\beta_0\|}\right].$$

For $r_n = n^{1/3}$, we have

$$n^{-1} r_n^4 E[M^2_{\frac{K}{r_n}}] = n^{-1} r_n^4 O(K r_n^{-1}) = O(1),$$

due to the assumption on the density of the angular component of $\boldsymbol{X}$. Hence, the first part of (C4) holds. $\forall \rho > 0$, $I\{r_n^2 M_{\frac{K}{r_n}}(\cdot) > \rho n\}$ is zero for all $n$ sufficiently large. Hence,

$$n^{-1} r_n^4 E[M^2_{\frac{K}{r_n}} I\{r_n^2 M_{\frac{K}{r_n}} > \rho n\}] \leq n^{-1} r_n^4 \frac{4}{\eta^2} P\{M_{\frac{K}{r_n}}(\cdot) > \rho r_n\} = o(1).$$

So the second part of (C4) also holds. By similar calculation, we can show (C5) holds.

By DVK's remark 3(iv), (C7) is fulfilled by defining the random function $W_n : \mathbb{R}^{(p-1)} \to \mathbb{R}$ as

$$W_n(t) = \langle \Gamma(\boldsymbol{\beta}_0, \hat{h}), t \rangle, \tag{14}$$

and define the deterministic bilinear function $V : \mathbb{R}^{(p-1)} \times \mathbb{R}^{(p-1)} \to \mathbb{R}$ as

$$V(t,t) = \frac{1}{2} t^T \Lambda(\beta_0, h_0) t, \tag{15}$$

where $\Gamma$ and $\Lambda$ are defined in the proof for Lemma 1.

Next we check (C9). In the proof of Theorem 1, we have shown $\Gamma(\widetilde{\boldsymbol{\beta}}_0, \hat{h}) = o_p(n^{-1/2+\gamma_0})$

10

for an arbitrarily small $\gamma_0 > 0$. Hence, for an arbitrary compact subset $\mathcal{K}$ of $\mathbb{R}^{p-1}$,

$$n^{1/3} \sup_{t \in \mathcal{K}, t \neq 0} \|W_n(t)\|t\|^{-1}\| = o_{P^*}(1).$$

Since $\Gamma(\widetilde{\boldsymbol{\beta}}_0, h_0) = 0$,

$$\begin{aligned}
&r_n^2 B_n\big(\boldsymbol{\beta}_0 + \big(0, r_n^{-1}t^T\big)^T, h_0\big) \\
&= n^{2/3} \big[\mathbb{P}_n g\big(\cdot, \boldsymbol{\beta}_0 + \big(0, n^{-1/3}t^T\big)^T, h_0\big) - \mathbb{P}_n g\big(\cdot, \boldsymbol{\beta}_0, h_0\big)\big] \\
&= n^{2/3} \Big[\mathbb{P}_n g\big(\cdot, \boldsymbol{\beta}_0 + \big(0, n^{-1/3}t^T\big)^T, h_0\big) - \mathbb{P}_n g\big(\cdot, \boldsymbol{\beta}_0, h_0\big) \\
&\quad - P g\big(\cdot, \boldsymbol{\beta}_0 + \big(0, n^{-1/3}t^T\big)^T, h_0\big) + P g\big(\cdot, \boldsymbol{\beta}_0, h_0\big)\Big] \\
&\quad + \frac{1}{2} t^T \Lambda(\widetilde{\boldsymbol{\beta}}, h)|_{\widetilde{\boldsymbol{\beta}} = \widetilde{\boldsymbol{\beta}}_0, h = h_0} t + o(1),
\end{aligned} \tag{16}$$

where the exact form of $\Lambda$ depends on whether $\beta_{01} = 1$ or $\beta_{01} = -1$, as discussed in the proof of Lemma 1. Therefore, the deterministic continuous function $\Psi(t)$ required in (C9) can be set to

$$\Psi(t) = \frac{1}{2} t^T \Lambda(\widetilde{\boldsymbol{\beta}}_0, h_0) t. \tag{17}$$

It remains to derive the limiting process of the first term on the right side of (16). Let $\mathbb{G}_n = \sqrt{n}(\mathbb{P}_n - P)$. Then the first term on the right side of (16) simplifies to

$$\begin{aligned}
&n^{2/3} \big(\mathbb{P}_n g\big[\cdot, \big\{\beta_{01}, \big(\widetilde{\boldsymbol{\beta}}_0 + n^{-1/3}t\big)^T\big\}^T, h_0\big] \\
&\quad - \mathbb{P}_n g\big(\cdot, \boldsymbol{\beta}_0, h_0\big) - P g\big[\cdot, \big\{\beta_{01}, \big(\widetilde{\boldsymbol{\beta}}_0 + n^{-1/3}t\big)^T\big\}^T, h_0\big] + P g\big(\cdot, \boldsymbol{\beta}_0, h_0\big)\big) \\
&= n^{1/6} \big\{\mathbb{G}_n \big[g\big(\cdot, \big\{\beta_{01}, \big(\widetilde{\boldsymbol{\beta}}_0 + n^{-1/3}t\big)^T\big\}^T, h_0\big) - g\big(\cdot, \boldsymbol{\beta}_0, h_0\big)\big]\big\}.
\end{aligned} \tag{18}$$

Note that this is the same as the centered process in the parametric case (see page 292 in Van Der Vaart and Wellner (1996)). The covariance function of the limiting process is

$$K(s_1, s_2) = \lim_{n \to \infty} n^{1/3} P\big\{g\big(\cdot, \big\{\beta_{01}, \big(\widetilde{\boldsymbol{\beta}}_0 + n^{-1/3}s_1\big)^T\big\}^T, h_0\big) g\big(\cdot, \big\{\beta_{01}, \big(\widetilde{\boldsymbol{\beta}}_0 + n^{-1/3}s_2\big)^T\big\}^T, h_0\big)\big\}. \tag{19}$$

Finally, the mean-zero process (18) converges weakly to a Gaussian process with covariance

function $K(s_1, s_2)$, which we denote by $\mathbb{W}(t)$.

To prove (C10), we can use similar techniques used in proving (B2) to prove $\mathcal{M}_\delta$ has finite uniform entropy integral, which is an alternative to the original condition in (C10) in terms of the bracketing integral (see (3.2.8) in Van Der Vaart and Wellner (1996)).

Summarizing the above, we have proved Conditions (C1)-(C10) of DVK hold. By Theorem 3 of DVK, $n^{1/3}(\widetilde{\boldsymbol{\beta}}_n - \widetilde{\boldsymbol{\beta}}_0)$ converges to the unique maximizer of the process $t \mapsto \Psi(t) + \mathbb{W}(t)$. $\square$

**Proof of Lemma 1**. Given an IDR $\boldsymbol{d} = \boldsymbol{d}_{\boldsymbol{\xi}} \in \mathcal{D}$, define

$$m_{\boldsymbol{\xi}}(b, X, D, Y, \Delta, G) = \frac{\widetilde{R}(\boldsymbol{d})\rho_\tau(Y-b)}{\widetilde{w}_{\boldsymbol{d}}},$$

where $\widetilde{R}(\boldsymbol{d}) = \Delta I(D_1 = d_{1,\boldsymbol{\beta}}(\boldsymbol{X}))\{I(Y \leq s) + I(Y > s)I\{D_2 = d_{2,\boldsymbol{\varsigma}}(H_2)\}\}$. Then $\widehat{Q}_\tau(\boldsymbol{\xi}, \widehat{G}_C)$ minimizes $M_n(b, \widehat{G}_C) = \mathbb{P}_n\{m_{\boldsymbol{\xi}}(b, X, D, Y, \Delta, \widehat{G}_C)\}$, where $\mathbb{P}_n$ stands for the empirical average.

Let $M(b, G_C) = E\{m_{\boldsymbol{\xi}}(b, X, D, Y, \Delta, G_C)\}$. we will first show that

$$M(b, G_C) = E(\rho_\tau(T^*(\boldsymbol{d}) - b). \tag{20}$$

Importantly, (20) implies that the marginal quantile $Q_\tau\{T^*(\boldsymbol{d}_{\boldsymbol{\xi}})\}$ minimizes $M(b, G_C)$. Note that (20) follows from two observations. First,

$$\begin{aligned}
& E\Big(\frac{\Delta I(D_1 = d_1(\boldsymbol{X}_1))I(Y \leq s)\rho_\tau(Y-b)}{\widetilde{w}_{\boldsymbol{d}}} \mid O^*(\boldsymbol{d})\Big) \\
=\ & E\Big(\frac{I(T \leq C)I(D_1 = d_1(\boldsymbol{X}_1))}{\pi_{D_1}(\boldsymbol{X}_1)G_C(Y)} I(Y \leq s)\rho_\tau(Y-b) \mid O^*(\boldsymbol{d})\Big) \\
=\ & E\Big(\frac{I(\dot{T}(d_1, \emptyset) \leq C)}{G_C(\dot{T}(d_1, \emptyset))} I(\dot{T}(d_1, \emptyset) \leq s)\rho_\tau(\dot{T}(d_1, \emptyset) - b) \mid O^*(\boldsymbol{d})\Big) \\
=\ & E\Big(I(\dot{T}(d_1, \emptyset) \leq s)\rho_\tau(\dot{T}(d_1, \emptyset) - b) \mid O^*(\boldsymbol{d})\Big)
\end{aligned}$$

.

12

Second,

$$E\Big(\frac{\Delta I\{D_1=d_1(\boldsymbol{X}_1)\}I\{D_2=d_2(\boldsymbol{X}_1,D_1,\boldsymbol{X}_2)\}I(Y>s)\rho_\tau(Y-b)}{\widetilde{w}_{\boldsymbol{d}}}\mid O^*(\boldsymbol{d})\Big)$$

$$= E\Big(\frac{I(T\le C)I\{D_1=d_1(\boldsymbol{X}_1)\}I\{D_2=d_2(\boldsymbol{X}_1,D_1,\boldsymbol{X}_2)\}}{\pi_{D_1}(\boldsymbol{X}_1)\pi_{D_2}(\boldsymbol{X}_1,D_1,\boldsymbol{X}_2)G_C(Y)}I(Y>s)\rho_\tau(Y-b)\mid O^*(\boldsymbol{d})\Big)$$

$$= E\Big(\frac{I(\dot{T}(d_1,d_2)\le C)}{G_C(\dot{T}(d_1,d_2))}I\big(\dot{T}(d_1,\emptyset)>s\big)\rho_\tau\big(\dot{T}(d_1,d_2)-b\big)\mid O^*(\boldsymbol{d})\Big)$$

$$= E\Big(I\big(\dot{T}(d_1,\emptyset)>s\big)\rho_\tau\big(\dot{T}(d_1,d_2)-b\big)\mid O^*(\boldsymbol{d})\Big).$$

Hence, (20) holds.

To prove that $\widehat{Q}_\tau(\boldsymbol{\xi},\widehat{G}_C)\to Q_\tau\{T^*(d_{\boldsymbol{\xi}})\}$ in probability, we will verify that the following three conditions are satisfied.

(i) $\sup\limits_{b<L}\big|M_n(b,\widehat{G}_C)-M(b,G_C)\big|\xrightarrow{P}0,$

(ii) $\forall\,\epsilon>0,\ \inf\big\{M(b;G_C):\big|b-Q_\tau\{T^*(d_{\boldsymbol{\xi}})\}\big|\ge\epsilon\big\}>M\big(Q_\tau\{T^*(d_{\boldsymbol{\xi}})\},G_C\big),$

(iii) $M_n\big(\widehat{Q}_\tau(\boldsymbol{\xi},\widehat{G}_C),\widehat{G}_C\big)\le M_n\big(Q_\tau\{T^*(d_{\boldsymbol{\xi}})\},\widehat{G}_C\big)+o_P(1).$

To prove (i), note that

$$M_n(b,\widehat{G}_C)-M(b,G_C)=\big[M_n(b,\widehat{G}_C)-M_n(b,G_C)\big]+\big[M_n(b,G_C)-M(b,G_C)\big]. \quad (21)$$

By the preservation theorems in §2.10.2 of Van Der Vaart and Wellner (1996), the class of functions $\mathcal{J}=\big\{m_b(X,A,Y,\Delta;G_C):b\in(0,L),\boldsymbol{\xi}\in\{-1,1\}\times\widetilde{\mathbb{B}}\times\{-1,1\}\times\widetilde{\mathbb{Z}}\big\}$ is Donsker. Hence,

$$\sup_{b<L}\big|M_n(b,G_C)-M(b,G_C)\big|\xrightarrow{a.s.}0. \quad (22)$$

By Csörgő and Horváth (1983), $\forall\,\gamma_0>0,\ \sup\limits_{t<L}\big|\widehat{G}_C(t)-G_C(t)\big|=o(n^{-1/2+\gamma_0})$ almost surely. Therefore, for $\forall\,\gamma_0>0,$

$$\sup_{b<L}\big|\boldsymbol{M}_n(b,\widehat{G}_C)-\boldsymbol{M}_n(b,G_C)\big|\le 2\sup_{t<L}\big|\widehat{G}_C(t)-G_C(t)\big|\Big|n^{-1}\sum_{i=1}^n\eta^{-2}\rho_\tau(Y_i-b)\Big|$$

$$=o(n^{-1/2+\gamma_0})O_p(1). \quad (23)$$

(23) together with (22) and (21) imply (i). Next, (ii) holds because $M(b, G_C)$ is continuous and uniquely minimized at $b = Q_\tau \{T^*(d_{\boldsymbol{\xi}})\}$. Lastly, (iii) is implied by the fact that the criterion function $M_n(\cdot, \widehat{G}_C)$ is minimized at $\widehat{Q}_\tau(\boldsymbol{\xi}, \widehat{G}_C)$.

We will next show the above properties lead to the consistency of $\widehat{Q}_\tau(\boldsymbol{\xi}, \widehat{G}_C)$. From (ii), for any small $\epsilon > 0$, $\exists\, \vartheta > 0$ such that for $b$ such that $|b - Q_\tau \{T^*(d_{\boldsymbol{\xi}})\}| \geq \epsilon$, we have $\boldsymbol{M}(b; G_C) > \boldsymbol{M}(Q_\tau \{T^*(d_{\boldsymbol{\xi}})\}; G_C) + \vartheta$. Thus

$$
\begin{aligned}
&P\Big(|\widehat{Q}_\tau(\boldsymbol{\xi}, \widehat{G}_C) - Q_\tau \{T^*(d_{\boldsymbol{\xi}})\}| \geq \epsilon\Big) \\
&\leq\ P\Big(\boldsymbol{M}(\widehat{Q}_\tau(\boldsymbol{\xi}, \widehat{G}_C); G_C) > \boldsymbol{M}(Q_\tau \{T^*(d_{\boldsymbol{\xi}})\}; G_C) + \vartheta\Big) \\
&\leq\ P\Big(\boldsymbol{M}(\widehat{Q}_\tau(\boldsymbol{\xi}, \widehat{G}_C); G_C) - \boldsymbol{M}_n(\widehat{Q}_\tau(\boldsymbol{\beta}, \widehat{G}_C); \widehat{G}_C) > \\
&\qquad\quad \boldsymbol{M}(Q_\tau \{T^*(d_{\boldsymbol{\xi}})\}; G_C) - \boldsymbol{M}_n(Q_\tau \{T^*(d_{\boldsymbol{\xi}})\}, \widehat{G}_C) + \vartheta/2\Big)(1 + o(1)) \\
&\leq\ P\Big(\sup_{b < L} |\boldsymbol{M}(b, G_C) - \boldsymbol{M}_n(b, \widehat{G}_C)| > \vartheta/4\Big)(1 + o(1)) \to 0,
\end{aligned}
$$

as $n \to \infty$, where the second inequality follows from property (iii) and the last step follows from property (i). Thus $\widehat{Q}_\tau(\boldsymbol{\xi}, \widehat{G}_C) \xrightarrow{P} Q_\tau \{T^*(d_{\boldsymbol{\xi}})\}$. $\square$

## 2 Additional numerical results

**Example S1 (comparing quantile criterion and mean criterion)** In this example, we compare the performance of the new method with the method of Simoneau et al. (2020) which considers a mean-optimal criterion. the method of Simoneau et al. (2020) is implemented using the function `DWSurv` in the R package `DTRreg` (see Wallace et al. (2020)). The goal is to illustrate that each has its own advantage depending on the specific target. Here, we generate the random data from a heteroscedastic regression model $\log(T) = (X_1 + X_2) + A(-X_1 + X_2)^3 + (0.5 + A*(0.5X_1 + 0.5X_2))\epsilon$ where $X_1$ and $X_2$ are independent random variables with Uniform$(0, 1)$ distribution, $\epsilon \sim N(0, 0.5)$ is independent of $\boldsymbol{X} = (X_1, X_2)^T$, and $A \sim$ Bernoulli$(0.5)$ is independent of $(\boldsymbol{X}, \epsilon)$. The censoring variable $C$ is generated from the Uniform$[0, 13]$ distribution. The censoring rate of the data is about 24%. Table 1 summarizes the simulation results for $n = 1000$ based on 300 simulation runs. We consider

the quantile-optimal IDR for $\tau = 0.25$ and $\tau = 0.5$, respectively. We evaluate the estimated 0.25 quantile ($\widehat{Q}_{0.25}$), 0.5 quantile ($\widehat{Q}_{0.5}$) and mean ($\widehat{\mu}_{\text{mean}}$) using a large independent sample of size $10^7$. The results suggest that the different optimal IDRs perform best with respect to their own targets.

|  | $\widehat{Q}_{0.25}$ | $\widehat{Q}_{0.5}$ | $\widehat{\mu}_{\text{mean}}$ |
|---|---|---|---|
| New ($\tau = 0.25$) | **2.220** | 3.045 | 3.298 |
| New ($\tau = 0.5$) | 1.883 | **3.290** | 3.427 |
| Simoneau et al. | 2.080 | 3.102 | **3.453** |

Table 1: Results for Example S1

**Example S2 (comparing the model-free method with model-based methods)** In general, different methods that target different optimality criteria lead to different optimal IDRs. In this example, we consider a model where different criteria yield the same optimal rule. In this setting, the estimated parameters corresponding to the optimal IDR are comparable. We focus on comparing the proposed model-free method with two model-based methods in two scenarios: (i) the outcome regression model is correctly specified; (ii) the outcome regression model is mis-specified. Specifically, we compare with the semiparametric method of Simoneau et al. (2020) and a model-based approach using censored quantile regression by Peng and Huang (2008). The approach of Simoneau et al. (2020) requires a model for the blip function (or knowledge of the treatment-covariate interaction effect). The approach of Peng and Huang (2008) targets the conditional quantile and requires to specify the full model for the outcome regression. In this example, the approach of Simoneau et al. (2020) fits a linear blip function, while the approach of the approach of Peng and Huang (2008) fits a censored quantile regression with the linear main effect and linear covariates and interaction effect. The method of Peng and Huang (2008) is implemented using the function `crq` in the R package `quantreg` (see Koenker et al. (2020)).

We first consider the setting of correct model misspecification. the log-transformed survival time is generated from $\log(T) = (-0.2 + 0.5X_1 - 0.5X_2) + 1.2A(0.2 - 0.6X_1 - 0.1X_2) + \epsilon$, where $X_1$ and $X_2$ are independent random variables with Uniform$(0,1)$ distribution, $\epsilon \sim N(0, 0.5)$ is independent of $\boldsymbol{X} = (X_1, X_2)^T$, and $A \sim$ Bernoulli$(0.5)$ is independent of $(\boldsymbol{X}, \epsilon)$. The censoring variable $C$ is generated from the Uniform$[0, 5]$ distribution. The censoring

15

rate of the data is about 29%. One can show that for the above location-scale model with i.i.d. random errors, the quantile-optimal treatment regime (no matter the quantile level $\tau$ of interest) is given by $I(0.2 - 0.6X_1 - 0.1X_2 > 0)$. This would also be the same optimal treatment regime for the expected mean optimal criterion.

Table 2 summarizes the bias and standard deviations for estimating the parameters indexing the optimal IDR, where the coefficient of $X_1$ is normalized to have absolute value 1, for $n = 1000$ based on 300 simulation runs. It also reports the mis-classificiation rate (MR), the average percentage of times the estimated optimal IDR does not match the theoretically optimal IDR. All three methods perform well in this scenario with the model-based methods having slightly smaller standard errors.

|  | Intercept | $X_2$ | MR |
|---|---|---|---|
| New ($\tau = 0.5$) | 0.006 | -0.027 | 0.009 |
|  | ( 0.102 ) | (0.150 ) |  |
| New ($\tau = 0.25$) | 0.009 | -0.036 | 0.011 |
|  | (0.095) | (0.151 ) |  |
| Simoneau et al. | -0.014 | 0.029 | 0.007 |
|  | (0.058) | (0.098) |  |
| Peng& Huang ($\tau = 0.5$) | -0.001 | 0.001 | 0.0005 |
|  | (0.064) | ( 0.114 ) |  |
| Peng& Huang ($\tau = 0.25$) | -0.001 | 0.004 | 0.001 |
|  | (0.073) | (0.116) |  |

Table 2: Results for Example S2 (correct model specification)

Next, we compare the performance of the three methods under model misspecification. The random data are generated the same as above except that $\log(T) = (-0.2 + 0.5X_1 - 0.5X_2) + 1.2A(0.2 - 0.1X_2 - 0.6X_1)^3 + \epsilon$. The censoring rate of the data is about 28%. Here the interaction effect is nonlinear but a monotone transformation of the linear index. The theoretically optimal IDR remains the same for all three methods. However, both Simoneau et al. (2020) and Peng and Huang (2008) fitted using linear models and hence are misspecified. Table 3 summarizes the results. We observe that the proposed new method still accurately estimates the optimal IDR while the other two methods both have considerable bias with inflated mis-classification rates.

**Example S3 (the case with more covariates)** We generate the survival time from the

16

|  | Intercept | x1 | x2 | MR |
|---|---|---|---|---|
| New ($\tau = 0.5$) | -0.012 | 0.000 | -0.009 | 0.016 |
|  | (0.148) | (0.000) | (0.160) |  |
| New ($\tau = 0.25$) | -0.005 | 0.000 | -0.026 | 0.018 |
|  | (0.152) | (0.000) | (0.172) |  |
| Simoneau et al. | -0.132 | 0.180 | -0.173 | 0.177 |
|  | (2.932) | (0.573) | (5.372) |  |
| Peng& Huang ($\tau = 0.5$) | -0.541 | 0.240 | 0.104 | 0.250 |
|  | (8.721) | (0.651) | (9.101) |  |
| Peng& Huang ($\tau = 0.25$) | -0.523 | 0.253 | 1.682 | 0.503 |
|  | (6.069) | (0.666) | (32.166) |  |

Table 3: Results for Example S2 (model mis-specification)

model $\log(T) = \boldsymbol{X}^T\boldsymbol{\alpha} + A(\boldsymbol{X}^T\boldsymbol{\beta}) + \epsilon$, where $\boldsymbol{X} = (1, X_1, \ldots, X_p)^T$ with the components being independent Uniform$(0, 1)$ random variables, $\epsilon \sim N(0, 0.5)$ is independent of $\boldsymbol{X}$, and $A \sim$ Bernoulli$(0.5)$ is independent of $(\boldsymbol{X}, \epsilon)$. The censoring variable $C$ is generated from the model $\log(C) = -0.7 + 0.8X_1 - 0.2X_2 + 0.6X_3 + 0.8X_4 + \epsilon$. We consider $p = 10$ and $20$.

- For $p = 10$, $\boldsymbol{\alpha} = (-0.2, 0.5, -0.5, 0.3, -0.4, 0.5, -0.2, 0.4, -0.1, 0, 0)$ and $\boldsymbol{\beta} = (0.5, -0.6, -0.1, 0.1, 0.5, 0, -0.6, 0.1, -0.6, 0.3, -0.3)$.

- For the $p = 20$, $\boldsymbol{\alpha} = (0.5, 0.6, -0.3, 0, 0, 0, 0, 0.4, 0, \ldots, 0)$ and $\boldsymbol{\beta} = (0.5, 1, 0.2, 0.3, -0.3, -0.2, -0.4, 0.5, -0.5, 0.4, 0.6, 0.5, -0.6, -0.5, 0.4 - 0.4, 0, 0.3, -0.2, 0.2, 0.3)$.

The censoring rate of the data is about 0.26 for 10 dimension; and about 0.28 for 20 dimension. Table 4 summarizes the average mis-classification rate (the average percentage of times the estimated optimal IDR does not match the theoretically optimal IDR) for dimensional 10 and 20. We observe that the genetic algorithm still yields sufficiently accurate estimation of the optimal IDR although the computation becomes substantially more intensive.

| $n$ | $p = 10$ | $p = 20$ |
|---|---|---|
| 500 | 0.066 | 0.028 |
| 1000 | 0.052 | 0.034 |
| 2000 | 0.043 | 0.046 |

Table 4: Average mis-classification rates for Example S3

17



\end{document}